\documentclass[fleqn,usenatbib]{mnras}

\usepackage[T1]{fontenc}
\usepackage{ae,aecompl}
\usepackage{soul}

\usepackage{multirow}
\usepackage{float}
\restylefloat{table}

\usepackage{amsmath}	
\usepackage{amssymb}	
\usepackage{float}
\usepackage[caption = false]{subfig}
\usepackage[final]{graphicx}

\usepackage[normalem]{ulem}
\usepackage{scrextend}
\usepackage{lineno}
\usepackage{CJK}

\title[Fate of AGN stars]{The impermanent fate of massive stars in AGN disks}

\author[Ali-Dib
\& Lin]{
Mohamad Ali-Dib$^{1}$\thanks{E-mail: malidib@nyu.edu} 
and Douglas N. C. Lin
\begin{CJK*}{UTF8}{gbsn}(林潮)\end{CJK*}$^{2,3}$ \\
$^{1}$Center for Astrophysics and Space Science (CASS), New York University Abu Dhabi, UAE\\
$^{2}$Department of Astronomy and Astrophysics, University of California, Santa Cruz, CA 95064, USA\\
$^{3}$Institute for Advanced Studies, Tsinghua University, Beijing, China
}

\date{Accepted XXX. Received YYY; in original form ZZZ}

\pubyear{2021}

\begin{document}
\label{firstpage}
\pagerange{\pageref{firstpage}--\pageref{lastpage}}
\maketitle

\begin{abstract}
Stars are likely to form or to be captured in AGN disks. Their mass reaches an equilibrium when their 
rate of accretion is balanced by that of wind.  If the exchanged gas is well mixed with the stellar core, 
this metabolic process would indefinitely sustain an ``immortal'' state on the main sequence (MS) and
pollute the disk with He byproducts. This theoretical extrapolation is inconsistent with the 
super-solar $\alpha$ element and Fe abundances inferred from the broad emission lines in active
AGNs with modest He concentration. We show this paradox can be resolved with a highly-efficient retention 
of the He ashes or the suppression of chemical blending.  The latter mechanism is robust in the geometrically-thin, 
dense, sub-pc regions of the disk where the embedded-stars' mass is limited by the gap-formation condition.
These stars contain a radiative zone between their mass-exchange stellar surface and the nuclear-burning core.  
Insulation of the core lead to the gradual decrease of its H fuel and the stars' equilibrium masses.  
These stars transition to their post-main-sequence (PostMS) tracks on a chemical evolution time scale 
of a few Myr.  Subsequently, the triple-$\alpha$ and $\alpha$-chain reactions generate $\alpha$ and Fe 
byproducts which are released into their natal disks.  These PostMS stars also undergo core collapse, 
set off type II supernova, and leave behind a few solar-mass residual black holes or neutron stars.
\end{abstract}




\section{Introduction}
\subsection{Background \& motivations}
\label{sec:motivation}
Studying active galactic nuclei (AGNs) is fundamental in understanding the formation and evolution 
of galaxies and their central supermassive black holes SMBHs\citep{kormendy2013}. These very 
bright objects are powered by the release of gravitational energy through 
viscous dissipation in gaseous accretion disks around SMBHs with masses $M_\bullet \sim 10^{4-9} M_\odot$
\citep{Lynden-Bell1969, rees1984}. 

 Based on optical and x-ray data \citep{soltan1982, 
fabian1999, elvis2002}, population
synthesis models \citep{yu2002, marconi2004, shankar2004, shankar2009} 
infer average values for the Eddington 
factor ($\lambda_\bullet \equiv 
L_{\rm AGN}/{L_{\rm E}}_{\bullet} \simeq 0.6$ {, although see \cite{raim} who proposed the existence of three distinct AGN populations with $\lambda_\bullet$ ranging from 0.05 to 0.6 as a function of their obscuration and $\epsilon_\bullet$}) and efficiency factor ($\epsilon_\bullet \equiv 
L_{\rm AGN} /{\dot M}_\bullet c^2 \simeq 0.06$) where $L_{\rm AGN}$, $L_\odot$, ${L_{\rm E}}_{\bullet} = 
M_\bullet {L_{\rm E}} _{\odot} /M_{\odot}$, and ${L_{\rm E}}_{\odot} = 3.2 \times 10^4 L_\odot$ 
are the AGN, solar, Eddington luminosity for SMBH and the Sun respectively.  The 
accretion rate ${\dot M}_\bullet \simeq 2 f_{\bullet} 
m_8\ {M_\odot {\rm yr}^{-1}}$ where $m_8 \equiv M_\bullet/10^8 M_\odot$ and $f_{\bullet} \equiv (\lambda_\bullet / 0.6) (0.06 / \epsilon_\bullet)$.

In a steady state, SMBHs' ${\dot M}_\bullet$ is fed by their surrounding disks with same 
accretion rate ${\dot M}_{\rm d}$.  The structure of these disks is determined by the
efficiency of angular momentum transfer\citep{lyndenbell1974, pringle1981}. With an
$\alpha_\nu$ prescription \citep{shakura1973} for viscosity ($\nu = \alpha_\nu c_{\rm s} H$ where
$\alpha_\nu$ is an efficiency factor, $\Omega=\sqrt{G M_\bullet /R^3} $ is 
the Keplerian frequency),
the radial distribution of
the gas surface density $\Sigma_{\rm g} = 2 \rho_{\rm c} H$, midplane density $\rho_{\rm c}$, 
temperature $T_{\rm c}$, sound speed $c_{\rm s}$, {disk thickness $H \simeq c_{\rm s}/\Omega$, }
aspect radio $h=H/R$, surface temperature 
$T_{\rm e}$ and radiative flux $Q^- = 2 \sigma T_{\rm e}^4$ can be determined\citep{frank2002} 
(for either dust $\kappa_{\rm dust}$
or electron scattering $\kappa_{\rm es}$) opacity  under the assumption 
of thermal equilibrium in which $Q^-$ is balanced by the rate of viscous 
dissipation $Q^+ _\nu = 9 \Sigma_{\rm g} \nu \Omega^2/4$.

With the conventional $\alpha_\nu$ prescription and the inferred 
${\dot M}_{\rm d} (M_\bullet)$, the outer regions
(at radii $R \equiv r_{\rm pc} \ {\rm pc} \gtrsim 10^{-2}$ pc) of
disks around SMBHs with $m_8 \sim 0.01-1$ are prone to gravitational 
instability GI \citep{paczynski1978} which leads to gravito-turbulence 
with $\alpha_\nu \sim 0.1-1$ \citep{lin1987, lin1988, zhu2010a, zhu2010b, 
kratter2008, martin2011, rafikov2015}. GI also leads to prolific 
rates of {\it in situ} star formation \citep{goodman2003, goodmantan2004, thompson2005} 
and to capture from the omnipresent nuclear clusters \citep{syer1991, artymowicz1993, davies2020}. 
Here we introduce and adopt a generic stellar evolution 
and pollution in AGN disks (SEPAD) model under the assumption that 
these processes lead to a self-regulated marginally-stable state 
with the gravitational $Q (\simeq {c_{\rm s} \Omega / \pi G \Sigma_{\rm g}} = 
{M_{\bullet} / {\sqrt 8} \pi \rho_{\rm c} R^3})$ value $\sim 1$
\citep{safronov1960, toomre1964}.  

For the SEPAD model, we assume the embedded stars provide an auxiliary power $Q^+ _\star (> Q^+ _\nu)$ 
generated by the conversion of hydrogen H into helium He during their main sequence MS 
evolution and that of He into $\alpha$ element along their post main sequence 
tracks. The deduced $h(R)$ (Eq. \ref{eq:hscale}) and $Q^- (R)$ \citep{goodman2003, thompson2005} 
for the modified thermal equilibrium (with $ Q^+ _\star \simeq Q^-$ 
in \S\ref{sec:diskcomp}) are consistent with those 
inferred from the reverberation mapping data \citep{starkey2022}, the 
microlensing light curve \citep{pooley2007, morgan2018, cornachione2020} and the observed 
spectral energy distribution (SED), 
especially over the infrared wavelength range\citep{sanders1989}.   
A direct implication of the SEPAD scenario is that embedded 
stars also return He, $\alpha$, and Fe byproducts to the AGN disks and these
pollutants are mixed with gas in the main-disk flow towards the SMBH.  Consequently,
the contaminants do not accumulate with the AGNs' evolution and the
disks' metallicity is independent of $z_\gamma$
\citep{artymowicz1993}.

In a recent work, \cite{cantiello1, jermyn2022} studied the evolution of 
stars embedded in an AGN disk with the \texttt{MESA} code \citep{mesa1,mesa2,
mesa3,mesa4,mesa5}.  Their simulations include both mass 
accretion and stellar wind loss with rates ${\dot M}_{\rm Bondi}$ 
and ${\dot M}_{\rm wind}$ respectively. They found that, within 
the inner few pc's, where the density in the disk midplane is
relatively high ($\rho_{\rm c} \gtrsim 6 \times 10^{-18}$g cm$^{-3}$), 
the Bondi accretion  
timescale ($\tau_{\rm B} \equiv M_\star/ {\dot M}_{\rm Bondi}$) is 
much shorter than the nuclear burning timescale.
Consequently, newly formed or trapped stars undergo fast mass growth.  
As their mass $M_\star (\equiv m_\star M_\odot)$ increases to 
$\sim 10^3 M_\odot$, their nuclear fusion (through the CNO cycle) generates 
a luminosity $L_\star \sim {L_{\rm E}}_\star \equiv m_\star {L_{\rm E}}_\odot$ (their Eddington 
limit), under which radiation pressure suppresses the accretion rate and 
intensifies the stellar wind.  \cite{cantiello1} further assume that the 
accretion and wind lead to inflow of pristine H-rich disk gas and outflow 
of He-rich byproducts, albeit the stars' mass growth is stalled when these 
processes reach an equilibrium.  In addition, they assume that the fresh H 
replenishment and He byproduct are fully mixed  throughout the stellar
interior, including their radiative zones (\S\ref{mixingsection}).  Under these two conditions, 
massive stars retain their mass $M_{\rm eq}$ and composition, and perpetually remain 
on the MS as ``immortal'' stars until the AGN phase is terminated with a severe depletion of 
the disk gas. In the stellar core, H is converted into He through the CNO cycle at a rate 
${\dot M}_{\rm He} \simeq L_\star/\epsilon_{\rm He} c^2$ where $\epsilon_{\rm He} \simeq 0.007$ is the 
H-to-He fusion efficiency. Multiple stars coexist to maintain a thermal equilibrium ($Q^- \simeq 
Q^+ _\star > > Q^+ _\nu $) with a surface density $s_\star \simeq Q^- / L_\star \simeq Q^- / m_{\rm eq}
L_{\rm E \odot}$.  The ``immortal'' stars' wind return their He ashes to the disk at the same rate 
(${\dot M}_{\rm He}$).  This feedback process increases the He surface density in the disk at an average 
rate up to ${\dot \Sigma}_{\rm He} = s_\star {\dot M}_{\rm He} = Q^- / \epsilon_{\rm H} c^2$ and the
He mass fraction in the disk gas to ${\it Y}_{\rm d} \simeq 2 \pi \int {\dot \Sigma}_{\rm He} R dR/ 
{\dot M}_{\rm d}$.   
Although the CNO burning process markedly increases the abundance ratio of N/(C+O) 
from that produced by the triple-$\alpha$ and $\alpha$-chain reactions during the PostMS evolution of 
some previous generation stars, it does not lead to changes in the  star's $\alpha$ (mostly C+N+O) and 
Fe mass fraction (${\rm Z}_\alpha$ and ${\rm Z}_{\rm Fe}$) from their pre-existing values, as long as the high local and 
global disk density is maintained. Consequently, the perpetual massive MS stars do not lead to increases 
in the ongoing-AGN disks's $\alpha$ and Fe mass fraction (${\mathcal Z}_{\alpha}$ and ${\mathcal Z}_{\rm Fe}$).
 
These predictions can be tested by the observational inferred metallicity of AGN discs.
With the aid of the \texttt{CLOUDY} models for emission-line physics, the magnitude of ${\it Y}_{\rm d}$, 
${\mathcal Z}_{\alpha}$, and  ${\mathcal Z}_{\rm Fe}$ of the disk gas
can be inferred from the ratios of the measured strength of various broad emission lines associated 
with AGNs \citep{osterbrock2006} at a range
cosmological red-shifts, $z_\gamma$.  An expository survey of spectroscopic data
associated with AGNs' broad line regions (BLRs)
\cite{2023MNRAS.tmp.2527H} indicates that a) ${\it Y}_{\rm d}$ is $\sim 30\%$ higher 
than its solar value \citep{osterbrock1982, shang2007}, b) $\alpha$ (C+N+O) element 
abundance ${\mathcal Z}_{\alpha}$ is 
$\sim 3-5$ that of the solar value \citep{nagao2006a, temple2021, lai2022}, 
c) N/(C+O) is larger than the solar value and increases with ${\mathcal Z}_{\alpha}$
\citep{hamann2002}, and d) Fe mass fraction ${\mathcal Z}_{\rm Fe}$ is comparable to its 
solar value\citep{dietrich2003, maiolino2003, wang2022}.

The modest ${\it Y}_{\rm d}$ elevation and the substantial ${\mathcal Z}_{\alpha}$ and ${\mathcal Z}_{\rm Fe}$
enhancements contradict the expected predominantly He 
yield from a population of {\it hypothetically ``immortal''} massive main sequence (MS) stars which do not evolve 
onto their post main sequence (PostMS) phase until either the disk is nearly hydrogen free with ${\it Y}_{\rm d} 
\sim {\mathcal O} (1)$ \citep{jermyn2022} or the stars' accretion rate is extensively suppressed 
by the severe depletion of the disk gas \citep{cantiello1}.  The former scenario is inconsistent with
the observed modest value of $Y_{\rm d}$ whereas the latter possibility is inconsistent 
with the observationally inferred super-solar ${\mathcal Z}_{\alpha}$ and nearly solar ${\mathcal Z}_{\rm Fe}$
for ongoing AGNs, independent of the cosmological redshift.

\subsection{Main goal: ending the MS}
In order to resolve the theoretical and observational tensions, we re-visit stellar evolution 
in the context of the SEPAD scenario. 
Although stellar-evolution models themselves do not depend on the SEPAD assumptions, they can 
provide useful constraints on the SEPAD models in terms of both the disk-environment boundary conditions and the implications 
on the He, $\alpha$ and Fe byproducts.
The primary motivation in this paper is to investigate some potential 
channels for the massive {\it evolving} stars to transition from the MS 
to PostMS phase.  This transition requires that the stellar hydrogen mass fraction 
$X_\star$ be severely depleted in the hydrogen burning zone.  Since the mass content of these embedded 
stars is recurrently reprocessed by accretion-wind metabolism, in contrast to the conventional standalone stars, 
it is essential to suppress the resupply of fresh hydrogen fuel into the nuclear furnace. 

In light that the modest $Y_{\rm d}$ inferred from observational data, we explore two potential interruption 
mechanisms for the hydrogen supply chain:

\noindent
$\bullet$ 1. an effective retention of the helium  byproducts carried 
by the stars' wind between their Bondi $R_{\rm B}$ and Hills $R_{\rm H}$ radii, or

\noindent
$\bullet$ 2. the isolation of the CNO burning core from the outer mass-exchange 
surface by a mixing-free radiative layer in the stellar interior.

\noindent
The second scenario is attainable provided a mass-growth limit ($\lesssim 6-700 M_\odot$)
is imposed by the suppression of stellar accretion rate due to the formation of tidally-induced 
gaps in the disk.

We carry out a series of numerical simulation with the \texttt{MESA} code
(\S\ref{sec:methods}). 
Results of these simulations are presented and analyzed in \S\ref{sec:endms}. 
We examine the feedback of the mass loss process on the chemical composition of the accreted 
material, evaluate its effectiveness in enabling the MS-to-PostMS transition and determine
the duration of the MS evolution $\tau_{\rm MS}$.  

\subsection{Secondary goals}
\subsubsection{Quantifying evolving stars' PostMS evolution}
The {\it evolving} stars' transition from MS to PostMS evolution (\S\ref{sec:PostMStrack}) not only 
activates the production of $\alpha$ elements through the triple-$\alpha$ 
and $\alpha$-chain reactions but also provides a robust channel for the 
{\it in situ} production of solar or supersolar ${\mathcal Z}_{\rm Fe}$ 
with subsolar Fe/$\alpha$ through (core-collapse) type II supernovae (SN II). 
In the hot cores of very metal-deficient (with  Log ${\rm Z}_\alpha/Z_\odot 
\lesssim  -1.75$ and presumably tenuous mass-loss rates), 
relatively massive ($\gtrsim 10^2 M_\odot$) PostMS stars, prolific pair 
production excites instabilities which trigger super novae and eject a large 
amount of $\alpha$ and Fe remnants without residual black holes \citep{woosley2017, spera2017}.
{These instabilities are suppressed by substantial mass losses in stand-alone
stars, although the stability metallicity threshold is mass and model dependent, with estimates ranging from log ${\rm Z}_\alpha/Z_\odot \gtrsim -1.75$ to $\gtrsim  -0.47$ \citep{pair1,pair4,pair3,pair2} }

As He is converted to $\alpha$ and Fe through triple-$\alpha$ and $\alpha$-chain reaction 
on their PostMS track, the magnitude of He mass fraction ${\rm Y}_{\star}$ inside the stars
reduces while both ${\rm Z}_\alpha$ and ${\rm Z}_{\rm Fe}$ increase. These byproducts 
are returned to the disk through PostMS winds and they contribute to the BLR's observed 
${\it Y}_{\rm d}$, ${\mathcal Z}_\alpha$ and ${\mathcal Z}_{\rm Fe}$.  The intensity 
of the stellar wind also determines the mass of the pre-collapse cores, the possibility 
of SNII, and the mass of remnant black-hole remnants rBHs. 
Along with SN II, accretion onto rBHs also contributes to $Q^+ _\star$. The
rBHs also provide seeds for intense VIRGO/LIGO gravitational wave events through mergers.  

A second goal of this investigation, with the aid of the \texttt{MESA} code (\S\ref{PostMSsection}),
is to estimate embedded PostMS stars' mass and luminosity evolution, the amount of He, $\alpha$,
and Fe yield returned to the disk, pre-collapse core mass $M_{\rm pc}$, their duration 
$\Delta \tau_{\rm PostMS}$, and stars' life span $\tau_\star$.  
Under the assumption of uninterrupted state of thermal equilibrium 
($Q^+ _\star \simeq Q^-$), we estimate the rate
of stars formation per unit surface area to be $\sim s_\star/\tau_\star$ and that of 
the heavy element pollution rate for the disk.  We also
set upper limits on the mass range of emerging rBHs after the SN II to be 
$M_{\rm pc}$. Since they are likely to be retained by the disk, these rBHs may gain substantial 
mass through subsequent gas accretion and multiple merger events.

\subsubsection{Multiple generations of evolving stars}

If embedded stars evolve through the MS-PostMS-SN II or collapse impermanent life cycle,
accumulative pollution from multiple generation of stars is also needed 
to elevate ${\it Y}_{\rm d}$, ${\mathcal Z}_{\alpha}$, and ${\mathcal Z}_{\rm Fe}$ 
to their observationally inferred values.  Moreover, these abundances are larger 
in the broad line region (BLR) than those in the narrow line region (NLR) 
\citep{nagao2006b, terao2022}. Their independence of $z_\gamma (\sim 0-7)$
\citep{nagao2006a, derosa2011, xu2018} suggests that the rich $\alpha$ and 
Fe contents produced by the embedded stars not only pollute the disks but are
also carried by the gas flow and consumed by the central SMBH. These inferences 
support the SEPAD's basic assumption of {\it in situ} chemical evolution (\S\ref{sec:motivation}).

Our final objective is hence to gather evidences for the {\it in situ} formation of multiple 
generation of evolving stars. Based on our numerical models and analysis, 
we show, in \S\ref{sec:secondgen},  that the enhanced N/(C+O) ratio implies secondary 
processes, i.e. the $\alpha$ (C+N+O)  production from the triple-$\alpha$ burning during 
the PostMS phase of an initial-generation  stars must be reprocessed through the CNO cycle 
during the MS stage of next-generation stars. Finally, we discuss some model uncertainties
in \S\ref{sec:numsimp} and summarize our results and discuss some observable
tests in \S\ref{sec:summary}.

\section{Methods}
\label{sec:methods}

\subsection{AGN disk structure}
\label{diskmodel}
The SEPAD model is constructed with
the widely-adopted $\alpha_\nu$-disk prescription \citep{shakura1973}
for effective viscosity $\nu (= \alpha_\nu c_s H$).  In a steady state, the mass flux in disk is given by:
\begin{equation}
    {\dot M}_{\rm d}  = 3 {\sqrt 2} (\alpha_\nu h^3/Q) M_{\bullet} 
\Omega  = {\dot M}_{\bullet}.
\label{eq:mdotdisc}
\end{equation}
\citep{goodman2003}.  With $Q \sim 1$ and $\alpha_\nu \sim 0.1-1$ 
for gravito-turbulence being marginally maintained, we find
\begin{equation}
\label{eq:hscale}
h \sim 0.019 {f_\bullet ^{1/3} r_{\rm pc}^{1/2} \over \alpha_\nu ^{1/3} m_8^{1/6}},
\ \ \ \ 
\Sigma_{\rm g} = 2 \rho_c H \simeq 90
{f_\bullet^{1/3} m_8^{5/6} \over \alpha_\nu ^{1/3} r_{\rm pc}^{3/2}}
\ {\rm g \over cm^{2}},
\end{equation}
\begin{equation}
\label{diskdeneq}
\rho_{\rm c} = 7.6 \times 10^{-16} {m_8 \over r_{\rm pc}^{3}} \ {\rm g \over cm^{3}},
\ \ \ \ 
c_{\rm s} 
\simeq 13 {f_\bullet^{1/3} m_8 ^{1/3} \over \alpha_\nu ^{1/3}} \ {\rm km \over s}.
\end{equation}

The magnitude of $h$ can be observational measured from the
reverberation mappings of fluctuating lampposts within a few 
gravitational radii $\textit{R}_\bullet\equiv GM_\bullet/c^2$.
In exposed outer regions of the disk, the aspect ratio of the 
photo-surface for the reprocessed radiation is typically a few 
$h$ \citep{garaud2007} and that measured ($\sim 10^{-2}$) for 
NGC 5548 \citep{starkey2022} at $r_{\rm pc} \sim 10^{-2}$ 
(a few light days) is consistent with Equation (\ref{eq:hscale}).

\subsection{MESA numerical model}
\label{sec:mesa}

For all of our simulations we use the same modified \texttt{MESA} code used and provided by \cite{cantiello1}. 
We hence define the classic Bondi accretion rate as:
\begin{equation}
\dot{M}_{\mathrm{Bondi}}=4 \pi R_{\mathrm{B}}^{2} \rho_{\rm c} c_{\rm s}.
\label{eq:bondiaccrate}
\end{equation}
This estimate does not include the effect of stars' tidal interaction with the
disk gas which leads to gap formation and reduction of the accretion rate (see further discussion
in \S\ref{sec:gaplimit} and \S\ref{sec:modroche}).

\begin{figure*}
\begin{centering}
        \includegraphics[scale=0.50]{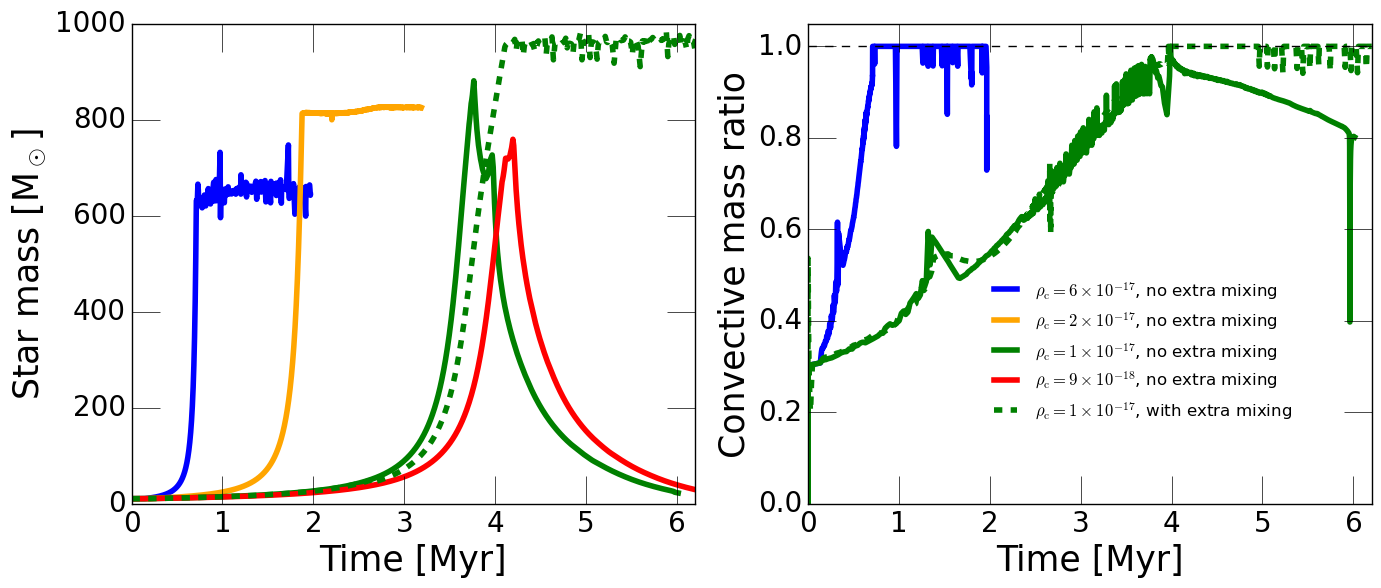}%
   \caption{{Left hand panel:} the star's mass evolution for different AGN disk densities. 
   In blue is our $\rho_{\rm c}=6\times 10^{-17}$ g/cm$^3$ nominal model with no extra mixing in the radiative zone. 
   Here the star reaches a ``immortal'' high-mass MS equilibrium state. For $\rho_{\rm c}=10^{-17}$ g/cm$^3$ 
   (solid green) however, also assuming no extra mixing, the star reaches a maximum mass of $\sim 900$ M$_\odot$ before 
   losing back most of it down to $\sim 20$ M$_\odot$ and evolving into PostMS stage. For the same disk density, if 
   we allow for extra mixing in the radiative zone similar to Cantiello et al. (2021) (dashed green), the star instead reaches 
   a ``immortal'' state. This indicate that extra mixing is irrelevant above a certain disk density, and controls 
   the star's evolution below it. We also show the evolutionary tracks for other densities with no extra mixing to delimit the fully convective star threshold value, found to be $2\times 10^{-17}$ g/cm$^3$.  {Right hand panel:} the evolution of the star's convective to total mass 
   ratio. A value less than 1 indicates the presence of a radiative zone. Considering the two cases without extra 
   mixing, we see that for $\rho_{\rm c}=6\times 10^{-17}$ g/cm$^3$ the star becomes fully convective, allowing it to efficiently transport material from the disk to the core, thus maintaining its equilibrium state. For $\rho_{\rm c}=10^{-17}$ g/cm$^3$ however, the star maintains a radiative zone, explaining its different evolutionary path. Adding extra mixing (dashed green curve) allows the star to be fully convective when close to the Eddington limit.}
    \label{fig:diffusion}
    \end{centering}
\end{figure*}

For trans-Eddington luminosities, the wind suppresses accretion resulting in a reduced Bondi mass-accretion 
rate that can be written parametrically as:
\begin{equation}
\dot{M}_{\mathrm{Bondi}, \Gamma}=\dot{M}_{\mathrm{Bondi}}\left(1-\tanh \left|L_{*} / {L_{\rm E}}_\star
\right|\right)
\label{eq:mdotbondigamma}
\end{equation}
where $L_\star$ and ${L_{\rm E}}_\star (= m_\star {L_{\rm E}}_\odot)$ are the stellar luminosity and its Eddington limit.
We adopt the same ``smoothing'' prescriptions as in \cite{cantiello1} for comparison purpose and it is an approximation of 
the modified ${\dot M}_{\rm Bondi}$ of Eq. \ref{eq:bondiacc} in \S\ref{sec:modbondi}.
The mass loss rate is also given parametrically by :
\begin{equation}
\label{eq:masslosse}
\dot{M}_{\mathrm{wind, \Gamma}}
=-\frac{L_\star}{V_\star^{2}}
\left[ 1+\tanh \left( 
\frac{L_\star-{L_{\rm E}}_\star} {0.1 {L_{\rm E}}_\star} 
\right) \right]
\end{equation}
with an escape velocity $V_\star = 
(2GM_\star/R_\star)^{1/2}$ from the stellar surface at radius $R_\star$ (this prescription
is equivalent to the analytic approximation for ${\dot M}_{\rm wind}$ in Eq. \ref{eq:windmod0}
\S\ref{sec:gap}).
Using these prescriptions and by setting the AGN density to $\rho_{\rm c}=6\times 10^{-17}$ g/cm$^3$ 
and sound speed to $ c_{\rm s}= 10^6$ cm/s, the star initially grows from $1 M_\odot$ to an 
accretion-wind equilibrium state with an asymptotic equilibrium mass $M_{\rm eq} \sim 630 M_\odot$.

We find that the AGN density parameter does not affect 
our results qualitatively as long as it is in the ``runaway accretion'' regime, reported by 
\cite{cantiello1} to be $\rho_{\rm c} \geq 6\times 10^{-18}$ g/cm$^3$.  
In a disk with marginal stability (i.e. $Q \sim 1$) around $10^8 M_\odot$ SMBHs, this
density includes all the region $\lesssim 2.3 m_8 ^{1/3}$ pc (Eq. \ref{diskdeneq}).
In contrast, typical radial extent of the BLR inferred from reverberation mapping data
gives a BLR radius $\sim 10-40$ light days \citep{bentz2013, horne2021}. 
However, the formation of gaps in the geometrically-thin, dense (Eqs. \ref{eq:hscale} \& 
\ref{diskdeneq}) sub-pc disk region near embedded stars' orbits may severely quench the gas supply
from the disk and impose growth limits on their masses (see further discussion
in \S\ref{sec:gaplimit} and \S\ref{sec:modroche}).

Throughout this paper, we follow \cite{cantiello1} in assuming that the Eddington luminosity is  given 
by ${L_{\rm E}}_\star = m_\star  {L_{\rm E}}_{\odot}$, hence ignoring its dependency on the hydrogen 
abundance through the electron-scattering opacity: ${L_{\rm E}}_\star = 4\pi GM_\star c/(0.2\,(1 + X_\star))$. 
This approximation does not introduce any significant qualitative effects on our results when we reran the 
simulations using the more self-consistent opacity definition.

Finally, we emphasize that all of our simulations were done with solar metallicity and 
composition for the accreted disk gas, in contrast with \cite{cantiello1} who used metal-free 
gas. The importance of this assumption is discussed in \S\ref{PostMSsection}.

\section{Transition from MS to PostMS tracks for the evolving stars }
\label{sec:endms}

\subsection{Isolating the stellar core}
\label{mixingsection}

In \cite{cantiello1}, mixing throughout the radiative zone (if it is
present) is imposed under the assumption that its efficiency equals  
to that in the convective region. This prescription was justified 
by noting that with the density and temperature distributions being nearly 
those of a $\gamma=4/3$ polytrope, the stars' internal structure
is prone to rotational instability and meridional circulation.  
This extra mixing in principle can allow the freshly accreted material 
to cross the radiative zone and to reach the core.  It also enables the 
nuclear byproducts in the core to be dredged up to their outer envelope.  
Without this extra blending, the radiative zone becomes a buffer which insulates 
the nuclear furnace in the stellar core from the (accretion-wind) mass-exchange 
zone near the stellar surface.  Here we relax this assumption and investigate 
the role of this parameterized mixing, in interrupting the supply of hydrogen 
to the core and enabling the MS-to-PostMS transition.

\begin{figure*}
\begin{centering}
        \includegraphics[scale=0.3]{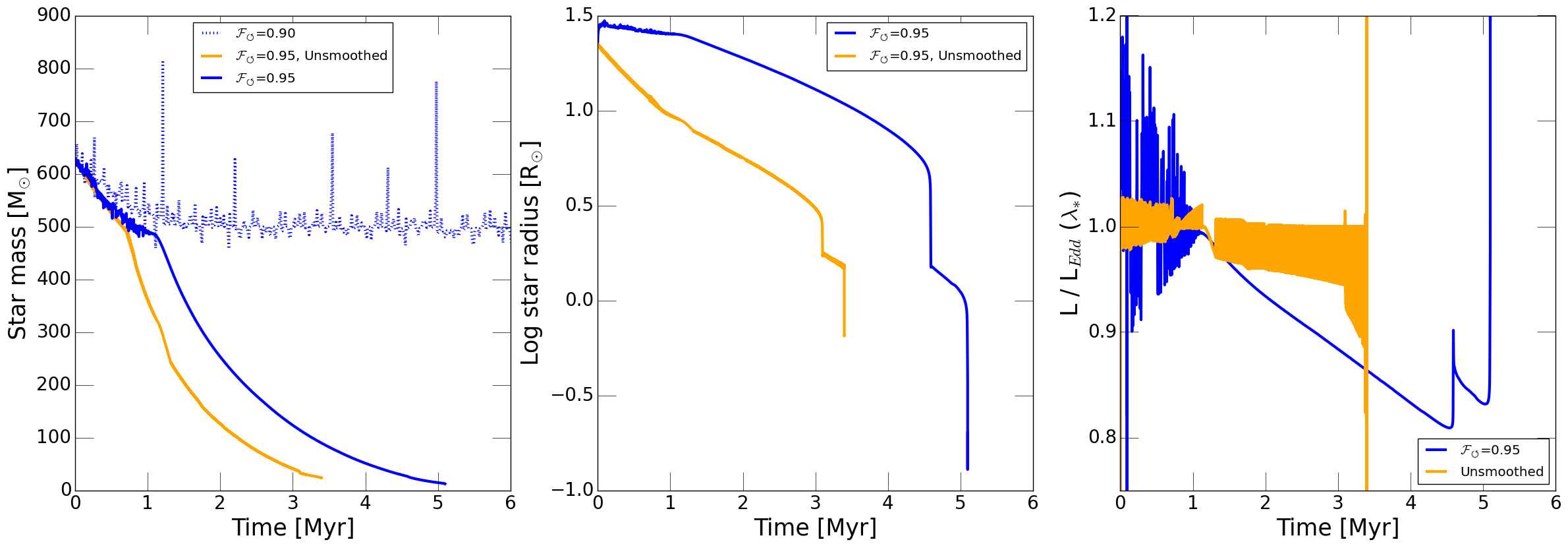}
   \caption{{left panel:} the solid blue line is our nominal model's star mass evolution during the mass loss phase after local wind-reaccretion been turned on, for $f_\circlearrowleft =0.95$. The orange line is the same case, but with accretion and mass loss numerical smoothing (\textit{tanh} terms in eqs. \ref{eq:masslosse} and \ref{eq:mdotbondigamma}) turned off, and the model changed so that the star cannot both accrete and lose mass during the same timestep. The dashed blue curve represents a case with $f_\circlearrowleft =0.90$, where wind-reaccretion is not efficient enough to force the star off the MS.
   {Central panel:} the stellar radii evolution curves for the same two $f_\circlearrowleft = 0.95$ cases of the left panel. 
   {Right panel:} the star's Eddington factor evolution for the same ($f_\circlearrowleft = 0.95$) nominal and unsmoothed models during the mass loss phases. The spikes at 4.6 and 5.1 Myr in the blue curve are due to post-main-sequence helium and carbon burning. 
   }
    \label{fig:main}
    \end{centering}
\end{figure*}

In Figure \ref{fig:diffusion} we show the evolution tracks for models starting from 1 M$_\odot$ star
with a solar composition, for different AGN disk densities. In all but one case, the radiative zone mixing has been turned off, 
allowing only for mixing in the convective zone. For the nominal ($\rho_{\rm c} = 6\times 10^{-17}$ 
g/cm$^3$) model, the star's evolution is nearly identical to the runaway cases in \cite{cantiello1}. This 
star quickly gains mass on a timescale $\lesssim 1$ Myrs and reaches a steady state with $M_\star \sim$ 
630 M$_\odot$ and the radius ${\rm R}_\star \sim$ 21.5 R$_\odot$. The stellar luminosity 
$L_\star$ increases from 1 L$_\odot$ to 10$^{7.5}$ L$_\odot$.
This asymptotic $L_\star$ is the Eddington luminosity of a star with this mass (${L_{\rm E}}_\star 
= m_\star {L_{\rm E}}_\odot)$, and hence this state represents the star's equilibrium point.
 
A slightly larger steady-state value of $M_{\rm eq} (\sim 810 M_\odot$) is obtained in 
the $\rho_{\rm c}= 2 \times 10^{-17}$ g/cm$^3$ model. This difference is mostly due to 
the difference in the boundary condition near the surface of the accreting stars.  
Fractional $M_{\rm eq}$-deviation between these and previous \citep{cantiello1} models 
is due to the solar versus metal-free composition adopted for the disk gas.  

In a lower-$\rho_{\rm c}$ ($1\times 10^{-17}$ g/cm$^3$) model, the star grows to 
$\sim 900 M_\odot$, before the emergence of a radiative zone.  The suppression
of mixing through this insulating layer leads to He enrichment in the core and net
mass loss over a timescale of few Myrs.  Eventually, the stars' mass is reduced to
$\lesssim 30 M_\odot$ as they undergo a MS-to-PostMS transition.  
This evolutionary pathway is physically distinct from the one found by \cite{cantiello1} 
for $\rho_{\rm c}= 5 \times 10^{-18}$ g/cm$^3$.  Transition in our model is caused by the 
state of mixing in the radiative zone instead of the competition between H-to-He conversion 
and H replenishment rates in the core. We furthermore find that, for solar-metallicity  gas, 
the area where extra mixing is relevant is limited to $\rho_{\rm c}= 6 \times 10^{-18}$ 
- $1 \times 10^{-17}$ g/cm$^3$ or over a radial extent $\sim 2 m_8 ^{1/3}$ pc in 
gap-free AGN disks. This domain is much further away from the SMBH than the region 
at 10-40 light days where the high $\alpha$-elements abundance is inferred from 
the broad-line ratios. For densities lower than $6 \times 10^{-18}$ g/cm$^3$, even 
with extra mixing, stars evolve into PostMS over $\sim 10^{8}$ yr as found 
by \cite{cantiello1}. 


To understand the origins of this mixing-driven evolutionary divergence between the 
models of immortal versus evolving stars, 
we inspect the ratio of the star's convective to total masses in Figure
\ref{fig:diffusion}, where we notice that while for the high density case the star 
is fully convective when it reaches the equilibrium state (${\dot M}_{\rm Bondi, 
\Gamma}={\dot M}_{\rm wind, \Gamma}$ with $\lambda_\star \simeq 1$), the lower 
density case maintains a permanent (outer) radiative zone. Without extra mixing, this radiative buffer 
prevents the freshly accreted gas from being transported to the core that burns 
its residual hydrogen before entering the PostMS stage.

Comparing this case against one with identical disk density ($\rho_{\rm c} =  1\times 10^{-17}$ g/cm$^3$) 
but including radiative zone extra mixing (also Fig. \ref{fig:diffusion}), we see that the star 
reaches and maintain a high-mass ``immortal'' state, similar to the higher disk density without 
extra mixing case. This dichotomy confirms that mixing in the radiative zone is controlling the fate of the 
star in the low density case. Finally, Figure \ref{fig:diffusion} also shows that 
embedded stars are fully convective when they reach an equilibrium state in disk regions with
$\rho_{\rm c} \geq 2\times 10^{-17}$ g/cm$^3$.

In summary, with the possible absence of gas mixing mechanisms other 
than convection, stars born in regions of the AGN disk where $\rho_{\rm c} \leq 1\times 10^{-17}$ 
g/cm$^3$ (at $\gtrsim 3.3 m_8^{1/3}$ pc in unperturbed regions of the disk, Eq. \ref{diskdeneq} or 
in a gap where the disk gas is locally depleted, Eq. \ref{eq:gap} in \S\ref{sec:gap}) 
evolve beyond the MS track.  Since additional mixing mechanisms may be present in
massive, rapidly rotating stars, we need to quantitatively access the likelihood
of rapid rotation for the massive stars and determine the efficiency of compositional 
blending under such circumstances. 


\begin{figure*}
\begin{centering}
        \includegraphics[scale=0.45]{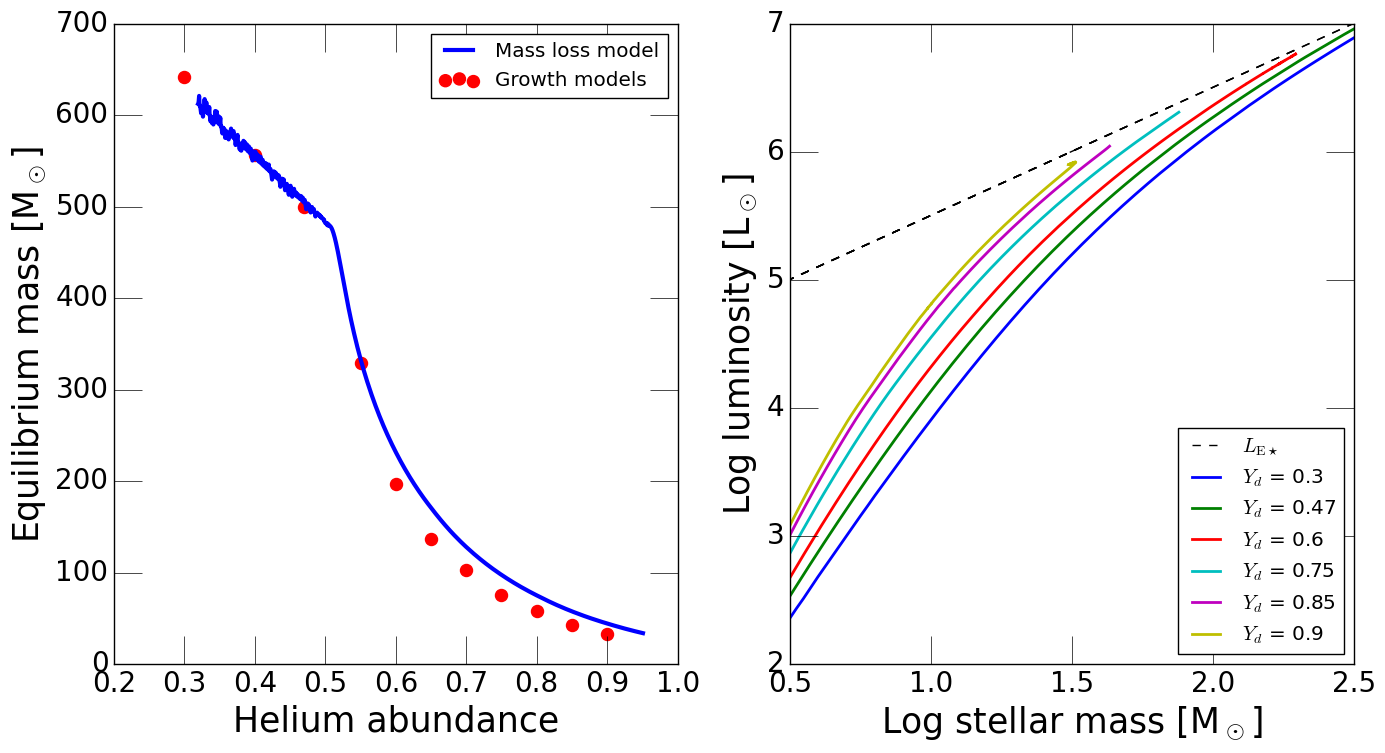}
          \caption{{Right hand panel:} Each solid line shows the evolution of the luminosity of embedded stars
          as they grow in mass (starting with $1 M_\odot$) through accretion from disks with a range of helium abundance 
          $Y_{\rm d}$ and the identical $\rho_{\rm c} (= 6 \times 10^{-17}$ g/cm$^3$). { The black dashed line is the Eddington luminosity $L_{\mathrm{Edd}}=3.2 \times 10^4\left({M_*}/{M_{\odot}}\right) L_{\odot}$ as function of stellar mass. As stated in section 2.2, we ignore the Eddington luminosity's dependency on the helium abundance, as it does not qualitatively affect our results.} 
          Stars with higher helium abundance intersect the Eddington luminosity and attain lower equilibrium masses.
          {Left hand panel:} the solid blue line is the star's mass evolution of our nominal case with high
          retention efficiency ($f_\circlearrowleft=0.95$).  This star's total helium mass fraction is plotted on
          the axis. The red scatter points are the asymptotic equilibrium masses of the growth models (where the color 
          solid lines for a range of $Y_{\rm d}$ intercept the black dashed line on the right hand panel). The similarity 
          between the line and dots indicates that the mass loss phase of our nominal model (with local retention) can be interpreted as continuous quasi-equilibrium states with increasing helium abundance and thus decreasing equilibrium mass. }
    \label{fig:helium}
    \end{centering}
\end{figure*}

\subsection{Re-accretion of He-byproducts in the wind.}
\label{sec:reaccrete}
\subsubsection{MESA implementation}
In this section we investigate the re-accretion of the He-byproducts ejected through 
the stars' wind as a possible mechanism for the MS-to-PostMS transition (\S\ref{sec:analyticrecycle}).

For illustration purpose, we adopt the nominal-$\rho_{\rm c}$ ($=6\times 10^{-17}$ g/cm$^3$) model with an initial 
mass $m_\star=1$.  Once the star's mass has reached an aymptotic equilibrium value, 
the model is saved using MESA's \texttt{save\_model\_when\_terminate} inlist flag. 
This stellar model is then loaded and resumed using a slightly modified code 
that incorporates an evolving chemical composition for the accreted material. 

Assuming that the accreted material is a mixture of wind-ejected and pristine-disk gas, 
the H mass fraction of the accreted material (controlled directly as a 
free parameter in \texttt{MESA}) becomes
\begin{equation}
\label{maineq}
X_{\rm acc} = f_{\circlearrowleft} {\rm X}_\star +(1-f_{\circlearrowleft} ) X_{\rm d}
\end{equation}
where ${\rm X}_\star$ is the H mass fraction in the star (assumed 
to be the same as that carried by the wind from of the star's surface), and $X_{\rm d}$ is that of the 
AGN disk (assumed to be its solar value). 
The retention efficiency $f_\circlearrowleft$ is implemented as a free model parameter in \texttt{MESA}
and $f_\circlearrowleft$ =0 or 1 for a totally open or closed system respectively.
Analogue prescriptions are also applied for helium and metals.
Finally, we let the star to evolve till either to a new equilibrium state is reached 
or to finish its evolution (we use MESA's default stopping condition when the 
core temperature reaches $10^{9.5}\ \mathrm{K}$).


Here we consider the case where significant fraction of the star's accreted material originates 
from locally-retained winds (inside the star's Roche radius, \S\ref{sec:modroche}).  
With $f_\circlearrowleft =0.95$ (Eq. \ref{maineq}),
we restart the simulation from the nominal equilibrium model with $M_\star = M_0 
= 630 M_\odot$ and $R_\star = 21.5 R_\odot$. Although the extra mixing through 
the radiative zone is switched off (\S\ref{mixingsection}), the star with this 
boundary condition ($\rho_{\rm c}=6\times 10^{-17}$ g/cm$^3$ ) and initial mass 
is fully convective and its internal chemical composition 
is fully homogenized between the outer mass-exchange zone and the nuclear-burning core.
In the absence of efficient retention (i.e. with $f_\circlearrowleft < < 1$), fresh H replenishment
would maintain this state indefinitely.

However, efficient retention and recycling of He increases the star's He mass fraction $Y_\star$
and molecular weight $\mu_\star$ over a timescale of $\sim$ 2-3Myr.  As the star adjusts to new 
hydrostatic quasi mass-exchange equilibria (with increasing $Y_\star$), it losses mass, at a fraction 
of $\dot{M}_{\mathrm{wind, \Gamma}}$ (Eq. \ref{eq:masslosse} and \S\ref{sec:equimassmu}). Its radius also shrinks
while its luminosity is maintained near its Eddington limit with $L_\star \simeq {L_{\rm E}}_\star$, 
(solid blue curves in the left, central, and right panels of Fig. \ref{fig:main}).  Over $\sim 3$ Myr, 
the star loses a substantial fraction of its initial mass on its MS track. When the star's interior 
eventually runs out of hydrogen, its evolution makes a transition from the MS to the PostMS phase.  


\subsubsection{Preserve a quasi accretion-wind equilibrium in chemically evolving stars 
through net mass loss.}
\label{sec:mstarmu}
Evolving MS stars with monotonically declining H fraction loss mass despite having established 
a quasi accretion-wind equilibrium with ${\dot M}_{\rm Bondi, \Gamma} \simeq 
{\dot M}_{\rm wind, \Gamma}$ and $\lambda_\star \simeq 1$.  The rate of net mass loss is 
a small fraction of ${\dot M}_{\rm wind, \Gamma}$.  To better understand the 
physical origins of this net mass loss along this MS evolutionary track, we plot (right 
panel of Fig. \ref{fig:helium}) the evolution of the stellar luminosity $L_\star$ 
during the growth of embedded star due to the accretion of 
disk gas with a range of $Y_{\rm d}$ (and thus $Y_\star$).  These stellar 
models are constructed with the same code and parameters 
(with no extra radiative zone mixing) as the growth phase of our nominal model.
Contributions from stellar wind is also included with the $f_\circlearrowleft=0$
open box prescription.   These growth tracks approaches the 
Eddington limit (dashed line) where the stars either attain an equilibrium mass 
and their mass growth is stalled in a ``immortal state''  
for $Y_{\rm d} \lesssim 0.5$, or start to loss mass and eventually evolve into 
PostMS for $Y_{\rm d} \gtrsim 0.5$. Note that \cite{jermyn2022} found a transitionary  $Y_{\rm d} \sim 0.9$, and this difference is due to the exclusion of extra mixing in our simulations.  

Figure \ref{fig:helium} shows clearly that higher helium abundance in the star ($Y_\star$) 
lead to lower Eddington-luminosity mass, and in consequence lower equilibrium  or gain-loss transition mass ($M_{\rm eq}$). 
The time evolution of our nominal simulation with wind-accretion 
(mass loss curve) is hence simply tracking this $Y_\star - M_{\rm eq}$ trend, 
as also shown on the left panel of the same plot. Physically, this is caused by the dependency of the 
luminosity-mass ($L_\star-M_\star$) relation on the mean molecular weight $\mu_\star$ 
of the star, and thus its helium abundance. 
For a constant mass $M_\star$, stars with increasing $\mu_\star$ have higher $L_\star$. 
Therefore, stars accrete from disk gas with different $Y_{\rm d}$ reach the Eddington luminosity 
${L_{\rm E}}_\star$ with different masses $M_\star$, leading to the $Y_{\star}-M_{\rm eq}$ relation we found in 
the left hand panel of the same plot. Analytic approximation for the quantitative $M_{\rm eq}$ dependence on the molecular 
weight $\mu_\star$ is presented in \S\ref{sec:equimassmu}.




We also show (right panel, Fig \ref{fig:main}) the $\lambda_\star \equiv L_\star/{L_{\rm E}}_\star$ controlling 
whether the star is accreting or losing mass. Overall, throughout the star's evolution, 
$\lambda_\star$ is close to unity as the star is simultaneously gaining 
and losing mass, although the trend is downwards as the star's luminosity decreases 
slightly steeper with mass than the Eddington luminosity. It is noticeable that while 
the luminosity is trans-Eddington, $\lambda_\star$ is actually less than 
1 for most of the simulation, even though the star is net losing mass. {This 
result is caused by the $tanh$ smoothing terms in equations \ref{eq:mdotbondigamma} and \ref{eq:masslosse} 
that allow the star to accrete mass for $\lambda_\star$ larger but close 
to 1, and to lose mass for $\lambda_\star$ smaller but close to 1. 
For $\lambda_\star =1$, $\dot{M}_{\rm wind,\Gamma}$/$\dot{M}_{\rm Bondi,\Gamma}$ 
= 0.84$\times$ $\dot{M}_{\rm wind}$/$\dot{M}_{\rm Bondi}$ instead of 0. On the other hand, 
for $\lambda_\star$ = 1.1, $\dot{M}_{\rm wind,\Gamma}$/$\dot{M}_{\rm Bondi,\Gamma}$ 
= 8.8$\times$ $\dot{M}_{\rm wind}$/$\dot{M}_{\rm Bondi}$. This smoothing prescription has no qualitative 
influence on our results, and is discussed further in \S\ref{sec:numsimp}. }

\begin{figure}
        \includegraphics[scale=0.45]{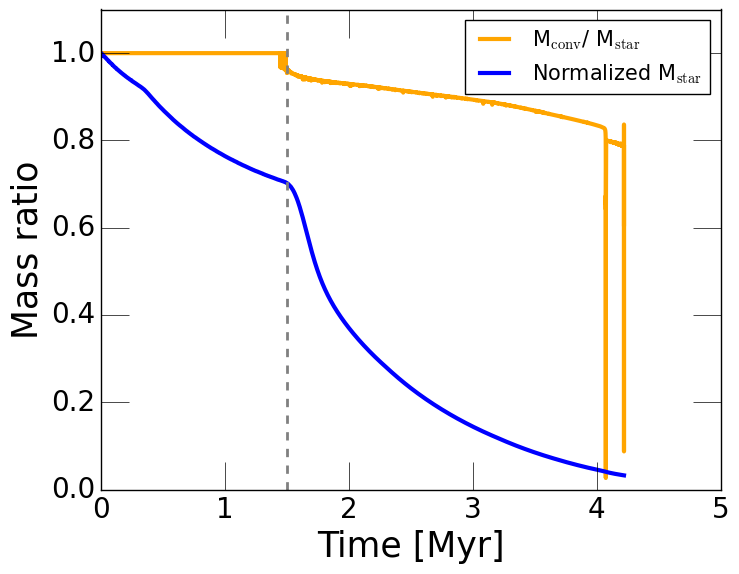}
   \caption{The blue solid curve is the time evolution of the star's normalized mass ($M_{\star}$/650 M$_\odot$) during the mass loss phase, and the yellow line is the convective core to full stellar mass ratio. The star's mass loss curve undergoes a slope change around 1.5 Myr that coincides with the opening of a radiative zone in the previously fully convective star.  } 
    \label{fig:main2a}
\end{figure}

\subsubsection{Compositional-mixing buffer by an outer radiative zone around a nuclear burning core.}
\label{sec:buffer}
An unexpected feature of the mass-loss curve is the change in slope around $m_\star \simeq 485$. 
To understand this change better we plot in Fig.~\ref{fig:main2a} the ratio of the star's total 
and convective core masses. The change in mass loss curve slope coincides 
with the expansion of an initially small outer radiative zone due to the gradual 
stellar-mass decrease and change in the opacity. The presence of this radiative zone 
influences the star's $M_\star$-$R_\star$ and $M_\star$-$L_\star$ relations (as functions
of $Y_\star$ and $\mu_\star$), changing the mass loss curve slope.
Moreover, it introduces a barrier for the freshly accreted material from the outer
mass-exchange region to reach the H-burning cores.  In the absence of 
extra mixing through the radiative zone, this insulating layer
leads to a close-box with $f_\circlearrowleft=1$ for stellar core (\S\ref{mixingsection}).
Thereafter, H in the nuclear furnace at the stellar core is exhausted on a time scale 
of $\Delta t_{\rm MS} \sim 2.5$ Myr (Fig. \ref{fig:main2a} and \S\ref{sec:equimassmu}) 
regulated by the chemical evolution (\S\ref{sec:equimassmu}) and 
the star undergoes a MS-to-PostMS transition when the H fuel in its core is exhausted. 

\subsubsection{Dependence on the retention efficiency.}
In Fig. \ref{fig:main} we also show the mass evolution curve for an 
identical simulation but with $f_\circlearrowleft$ = 0.90. In this case, 
while the star initially loses $\sim$ 100 M$_\odot$ of mass, it quickly 
re-establishes a new equilibrium around $m_\star= m_{\rm eq} \sim$ 500, $X_\star \sim 0.51$, 
and $Y_\star \sim 0.48$. The star remains fully convective such that the 
fractional amount of fresh supply from the disk gas can fully mix with
the nuclear burning core.  This modest amount of elemental metabolism enables the star
to become ``immortal'' and to stay on the main sequence.  This numerical result is consistent
with the inference from analytic approximation (Eq. \ref{eq:xdeclinecon}
and \S\ref{sec:analyticrecycle}).

The transitional stellar mass (below which there is a significant radiative zone) 
may vary slightly depends on $\mu_\star$, $\kappa$, and $\rho_{\rm c}$.  For discussion purpose 
below, we adopt the
value of the nominal model.  It is imperative to note that this critical mass is only comparable
to $M_{\rm eq}$ at tenuous gap-free outer ($r_{\rm pc} \gtrsim 5$ \S\ref{sec:mesa} and Fig. 
\ref{fig:diffusion}) region or $M_{\rm gap}$ due to gap formation in the inner dense, 
geometrically-thin regions of the disk (below).

\subsection{PostMS transition under growth limit due to disk gaps}
\label{sec:gaplimit}
In the previous subsection, we show that local retention and re-accretion of wind-ejected helium are
viable mechanisms to enable stars to undergo MS-to-PostMS transition in the tenuous (with sufficiently
low $\rho_c$) outer ($R \gtrsim$ a few pc) or in the dense inner disk region provided the retention 
factors $f_\circlearrowleft$ $\gtrsim$ 0.95. Although such a high retention efficiency 
may be challenging to accomplish, we note that it is only required for the fully convective
massive stars (\S\ref{sec:buffer}). Less massive stars 
have a radiative zone which can prevent the accreted disk gas from reaching the nuclear burning core and
mixing with the He-byproducts.  For stars embedded in regions with $\rho_{\rm c} = 6 \times 10^{-17}$g cm$^{-3}$,
radiative zone occurs in stars with $M_{\rm eq} \lesssim 485 M_\odot$ (Fig. \ref{fig:main2a}).  In regions with 
lower $\rho_{\rm c}$, the onset of radiative zone occurs in stars with larger $M_{\rm eq}$ 
(Fig. \ref{fig:diffusion}).  

\begin{figure}
        \includegraphics[scale=0.44]{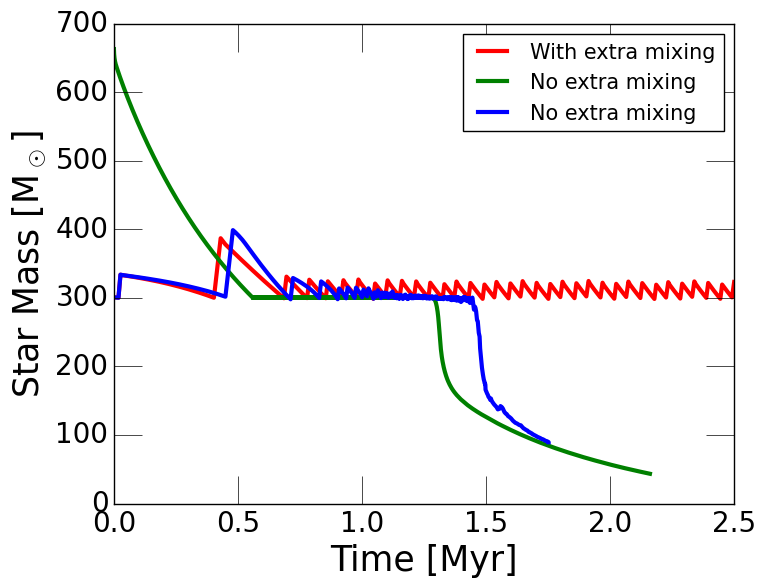}
   \caption{Simulations showing the effects of accretion suppression above 300 M$\odot$ (through gap opening), coupled 
   with the absence of extra mixing in the radiative zone. All simulations were done with $f_\circlearrowleft =0$. Stars starting at 300 M$\odot$ have an outer radiative zone and thus cannot maintain a steady state without extra mixing, and will evolve to postMS. Stars starting at 600 M$\odot$, even though fully convective at first, have the same fate.     } 
    \label{fig:gap}
\end{figure}

A natural physical process to stall accretion is the formation of gaps in the disk
around the embedded stars (\S\ref{sec:modroche}).  In the inner ($r_{\rm pc} < 1)$ dense, thin-disk region, gap-formation sets an upper growth limit to $M_{\rm gap} 
\lesssim$ a few $10^2 M_\odot$ (Eq. \ref{eq:gap} with $\lambda_\star < < 1$). 
{Note that the gap-formation and growth-stalling conditions are determined by
the disks' aspect ratio rather than by the unperturbed $\rho_{\rm c}$ (\S\ref{sec:modroche}) such that they can set  an upper limit for $M_\star$ over a wide radial range in the disk's sub-pc inner region.}
Nevertheless, gap formation decreases the ambient disk density which is also 
favorable for the preservation of the embedded stars' radiative zone (cf Figs. \ref{fig:diffusion} \& \ref{fig:main2a}). {We also note that as shown in eq.  \ref{eq:gap}, $M_{\rm gap}$ scales as $h^3$. As a nominal case we assumed $h \simeq 10^{-2}$ as discussed in section 2.1 based on the reverberation mapping of the metal-rich BLR 
in NGC 5548. In the outer regions of AGN disks (beyond a few pc), $h$ may be a few times higher, 
leading to the possibility of very high $M_{\rm gap}$'s. However, regions with such high $h$'s 
and marginal gravitational instabilities are likely to fragment and to break up into systems 
of clumpy clouds (similar to AGNs' dusty tori). In such porous medium, the stellar accretion
rate is substantially smaller than ${\dot M}_{\rm Bondi}$ (Eq. \ref{eq:bondiaccrate}).  (Moreover, 
in these regions, $\rho_{\rm c}$ is sufficiently small for radiative zone to 
be preserved in very massive stars, see \S\ref{mixingsection} and Fig. \ref{fig:diffusion}).}
Based on the results of \S\ref{sec:buffer}, we now consider the possibility 
that the accretion is quenched with $M_\star \lesssim M_1 \simeq 485 M_\odot$ so that 
the presence of radiative zone would lead mixing
insulation and MS-to-PostMS transition regardless the magnitude of $f_\circlearrowleft$ 
for the outer mass-exchange region.

For illustration, we present in Fig. \ref{fig:gap}, three models with the same parameters 
($\rho_{\rm c}=6\times 10^{-17}$ g/cm$^3$, sound speed to $ c_{\rm s}= 10^6$ cm/s,
$f_\circlearrowleft =0$) and accretion and wind rates (Eqs \ref{eq:mdotbondigamma} \& 
\ref{eq:masslosse}) as the nominal model.  We introduce a cutoff in the accretion 
rate for stars with $m_\star \geq 300$.
For the first and second models, we start with the nominal (with $m_\star =630$ and 
enhanced $Y_\star$) and a lower-initial-mass ($m_\star =300$ and solar composition) 
model respectively.  We do not include extra mixing in the 
radiative layer.  In both models, stars evolve onto
PostMS phase with similar evolutionary tracks, transitional and pre-core-collapse masses 
as those of the nominal model with $f_\circlearrowleft = 0.95$.  

We simulated another MESA model with an initial mass $m_\star=300$.  In this 
model, we include the extra mixing prescription for 
the radiative zone following the prescription by \cite{cantiello1}.  In this
case, the star evolves into a ``immortal'' equilibrium state with $m_\star \simeq 300$,
and $X_\star \simeq 0.5$.  Despite the enhancement in $Y_\star$, $\mu_\star$, 
$L_\star < < {L_{\rm E}}_\star$ (i.e. $\lambda_\star < < 1$).  This result is consistent
with our analytic approximation (\S\ref{sec:analyticrecycle} and \S\ref{sec:equimassmu}).
Transition from MS to PostMS occurs provided the freshly accreted disk gas does not replenish
and refresh the nuclear burning core.  These results clearly indicate the importance of the 
radiative buffer zone which prevents the replenishment of fresh supply of disk gas from
reaching the nuclear burning core.

\begin{figure*}
        \includegraphics[scale=0.48]{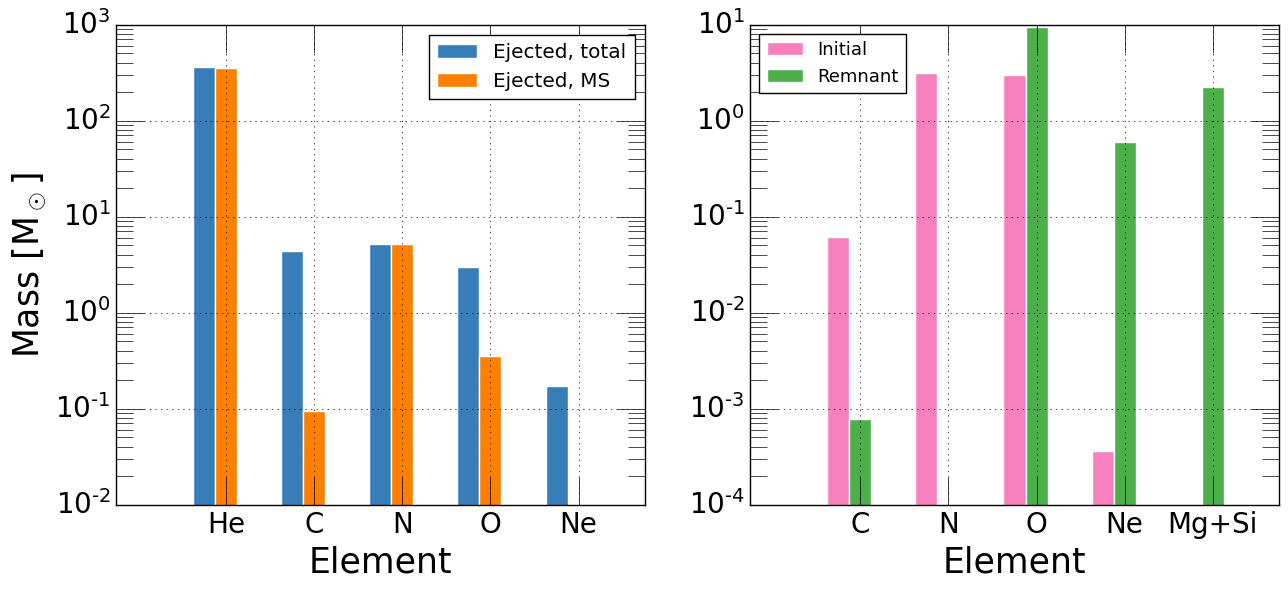}
   \caption{{Left panels:} a compositional breakdown of the total wind-ejected mass in our nominal local-retention model. Orange bars are the total masses ejected during the main sequence, while the blue bars are those ejected during the PostMS stage. {Right hand panel:} Green bars: the chemical composition of the remnant stellar object at the end of the simulation. Note that the remnant is at the onset of Si burning, and its mass can thus decrease further. Pink bars: the chemical composition of the star at the start of the simulation. } 
    \label{fig:yield}
\end{figure*}

\subsection{Stellar heating rate, population, and mass function.}
\label{sec:starpop}
In \S\ref{sec:motivation}, we indicated that the main motivation for the SEPAD model is to provide
adequate auxiliary power with a stellar surface number density $s_\star \simeq Q^-/L_\star$.  
For ``immortal'' stars, $L_\star \simeq m_{\rm eq} L_{\rm E \odot}$.  But since
the luminosity of {\it evolving} stars decreases  with time, $s_\star$ is determined by its time
averaged value (from the MESA model) over their life span $\tau_\star \simeq 5.1$ Myr,
\begin{equation}
    {\bar L}_\star = {1 \over \tau_\star} \int_0 ^{\tau_\star} L_\star dt \simeq {2.8 \times 10^{54} {\rm erg} 
    \over 5.1 {\rm Myr}} = 1.7 \times 10^{40} {\rm erg} {\rm s}^{-1}
\label{eq:averagelum}
\end{equation}
which is equivalent to the Eddington luminosity of a time-average ${\bar M}_{\rm eq} \simeq 143 M_\odot 
\simeq 0.23 M_0$
stars. In order to provide adequate auxiliary power to main marginal self gravity for the disk, 
the magnitude of $s_\star ({\bar M}_{\rm eq} )= Q^-/ {\bar L}_\star $ for this evolving 
population needs to be $\sim 4.4$ times larger than that, $s_\star (M_0 )
= Q^-/ L_\star (M_0) $, of the ``immortal'' stars with 
$M_0=630 M_\odot$, $X_0=0.69$, $Y_0=0.29$, and $Z_0=0.02$.  Since the $\alpha$ and Fe released
per individual embedded PostMS stars is approximately the same and the total amount of He
yield is determined by the required auxiliary power to maintain a state of marginal gravitational
stability, this enlargement of the stellar population ($s_\star$) 
enhances the disk-pollution rates of $\alpha$ and Fe relative to that of He (\S\ref{sec:productionalpha}).

Within a disk radius $R_{\rm d}$ (where constant-${\dot M}_{\rm d}$ flow can be approximately 
maintained), the total population of stars is $N_\star \simeq 2 \pi \int_0 ^{R_{\rm d}} 
s_\star R dR$ in this model.  In a steady state with a constant $N_\star$ or $s_\star$, 
the formation rate ${\dot N}_\star \simeq N_\star/\tau_\star$ is a constant.  The stellar 
mass function
\begin{equation}
    {m_\star \over N_\star} {d N_\star \over d m_\star} \simeq  {{\dot N}_\star \over N_\star} 
    {m_\star \over \vert {\dot m}_\star \vert}
    \simeq {1 \over \tau_\star (A_{\rm y} {\dot Y}_\star/Y_\star \vert_{\rm MS} + 
    A_{\rm z} {\dot Z}_\star/Z_\star \vert_{\rm PostMS}) }
    \label{eq:massfun}
\end{equation}
varies with $m_\star$ (rather than a single power law) which 
can be evaluated numerically (from the first part of 
Eq. \ref{eq:massfun} and the left panel of Fig. \ref{fig:main}) to be 
$\sim {m_\star / \vert {\dot m}_\star} \vert \tau_\star \sim 0.2$ for the
mass ranges $m_1 \geq m_\star \geq m_2$ and $m_2 \geq m_\star \geq m_3$ 
on the MS and PostPM tracks respectively (where $m_1$, $m_2$, and $m_3$ are defined 
in \S\ref{sec:diskcomp}). The mass function can also be 
analytically approximated with a similar value (from the second 
equation in Eq. \ref{eq:massfun} where the first and second terms 
on the denominator corresponds to MS and PostMS stars and
$A_{\rm y}$, $A_{\rm z}$, ${\dot Y}_\star (= -{\dot X}_\star)$, 
and ${\dot Z}_\star$
are estimated in \S\ref{sec:analyticrecycle} and
\ref{sec:equimassmu}).  
This top-heavy mass function 
is similar to the IMF inferred by \cite{miller1979}.

Based on this mass function, we can compute the average luminosity
\begin{equation}
    {\bar L}_\star = \int L_\star dN = \int_{m_3} ^{m_0} L_{\rm E \odot} m_\star {d N_\star \over d m_\star} d m_\star 
    \simeq 0.23 N_0 m_0 L_{\rm E \odot}
\end{equation}
for a population of quasi-equilibrium stars with $N_0$ as its normalization factor.
This total luminosity is equivalent to the Eddington luminosity of a time-average 
${\bar M}_{\rm eq} \simeq 143 M_\odot \simeq 0.23 M_0$ stars (Eq. \ref{eq:averagelum})
for the fiducial stellar model.  The equivalent mass ${\bar M}_{\rm eq}$ would be modified 
by $M_{\rm gap}$ if the stellar mass growth is quenched by gap formation (a smaller 
${\bar M}_{\rm eq}$ would lead to a larger $N_0$ and $s_\star$).
This mass function also has implications on the merger probability of coexisting stars 
(with $s_\star ({\bar M}_{\rm eq}) \sim 4.4 Q^-/m_0 L_{\rm E \odot}$ and total population
$N_\star$), star formation rate per unit area (${\dot s}_\star = s_\star / \tau_\star$), 
and over the entire disk ${\dot N}_\star$ which will be 
considered in future investigations.

\section{Effects on the composition of the  disk}
\label{sec:PostMStrack}
\subsection{MESA model of PostMS evolution}
\label{PostMSsection}
In this section we quantify the star's PostMS evolution and its effects on the chemical composition 
of the AGN disk. We hence investigate the star's chemical yield and the composition of the remnant. 
We focus on the local wind retention scenario, as we find it to be the more realistic mechanism to 
take the ``immortal'' stars off the MS. 
For our nominal model with $f_\circlearrowleft =0.95$, the MS-PostMS  transition
occurs and He burning starts after 4.7 Myr. This corresponds to the first sharp 
peak in  the $\lambda_\star \equiv L_\star/{L_{\rm E}}_\star$ curves of Fig. \ref{fig:main}. 
The second peak, at 5.1 Myr, corresponds to the onset of Ne burning and the end of the 
simulation  as the star's center reach a temperature of $10^{9.5}\ \mathrm{K}$. 
Above this temperature, our 21 elements nuclear network becomes insufficient. At this 
point the star is $\sim$ 13 M$_\odot$ and it is expected that silicon burning will follow 
till exhaustion, leading to a possible core collapse and type II supernovae explosion.

\subsection{Disk's He pollution rate by evolving and immortal stars}
\label{sec:diskcomp}
In the limit of high retention efficiency ($f_\circlearrowleft \gtrsim 0.95$), 
negligible He, $\alpha$ and Fe-byproducts are released to the disk and reaccreted onto the stars
through stellar metabolism.  Nevertheless, there would be a net loss of He 
from the stars to the disk if $M_{\rm eq}$ declines due to the accumulation of He 
and the increases in both $Y_\star$ and $\mu_\star$.

In the fiducial model with $f_\circlearrowleft=0.95$, stars' evolution can be characterized by
three stages (\S\ref{sec:equimassmu}).  Stage 1: the radiative region of high 
masses from $M_\star \sim M_0$
(with $Y_0 = 0.29$) to $M_1 = 485 M_\odot$ (with $Y_\star=Y_1 \simeq 0.5$) 
and their $M_\star-Y_\star$ dependence can be approximated by $A_\mu 
\equiv d {\rm ln} M_{\rm eq} / \partial {\rm ln} \mu_\star 
\simeq -2$ (Eq. \ref{eq:dlnmdlnmu}). 
Stage 2: MS stars loss mass between $M_1 \leq M_\star \leq M_2 \simeq 28 M_\odot$ 
(with $Y_2 =1$ and $A_\mu \sim -4.7$) and an expanding 
outer radiative zone which insulate stars' nuclear burning core from 
their stellar surface and impose $f_\circlearrowleft=1$. When the residual H in the core
is exhausted, they evolve onto the PostMS track. Stage 3: PostMS stars continue to 
reduce their mass from $M_2$ to $M_3 \simeq 13 M_\odot$ as they convert He into 
mainly $\alpha$ elements with an efficiency $\epsilon_\alpha \simeq 
10^{-3}$.  Once again, the increase in $\mu$ leads to a decrease in 
the accretion-wind equilibrium mass. Soon after C burning is initiated.  
the remaining cores collapse and undergo SNII with $\alpha$, Fe ejecta 
and rBHs.

As He is produced during the main sequence (stages 1 and 2), their release
rate into the disk is ${\dot M}_{\rm He}=- Y_\star {\dot M}_{\rm eq}$ 
(Eq. \ref{eq:dlnmdlnmu}) where ${\dot M}_{\rm eq} = A_{\rm y} M_{\rm eq} {\dot Y}_\star 
/ Y_\star$ is the changing rate of the equilibrium mass $m_{\rm eq} (Y_\star) 
\propto Y_\star ^{A_{\rm y}}$ (Eq. \ref{eq:mux0z1}).  
Neglecting variations in $A_{\rm y}$, the total He return to the disk is
\begin{equation}
    \Delta M_{\rm He} = -\int_{Y_{1}} ^1  A_{\rm y} M_{\rm eq} (Y_\star)  d Y_\star 
    \sim  {(Y_{\rm 1} M_1 - M_2) A_{\rm y} \over (1+ A_{\rm y})} \simeq 340 M_\odot
    \label{eq:hereturn}
\end{equation}
which is comparable to the He yield from the MESA model ($355 M_\odot$, left panel, Fig. \ref{fig:yield}).

During MS stages 1 and 2, $\Delta M_{\rm H} = 275.4 M_\odot$ of hydrogen was not burned and was returned to the disk.
This amount is slightly larger than the difference between the star's initial mass ($\sim 
630 M_\odot$, \S\ref{mixingsection}) and $\Delta M_{\rm He} + M_2 (\simeq 383)$.  This small discrepancy 
can be attributed to the much reduced but still finite accretion of disk gas during the MS evolution.
Subtracting the initial He mass (with initial $Y_0 = 0.29$, $Z_\star =Z_0 =0.02$ and $M_\star=630 M_\odot$), 
the total change $\Delta Y \simeq \Delta M_{\rm He} (1- Z_\star)/(\Delta M_{\rm H}+ \Delta M_{\rm He}) 
- Y_0 \simeq 0.26$.

Averaged over the stellar lifespan $\tau_\star \sim 4-5$ Myr (Fig. \ref{fig:main} \& \S\ref{sec:equimassmu}), 
the He pollution rate in the disk is ${\dot M}_{\rm He} \sim ({\Delta M}_{\rm He}- M_0 \times Y_0)
/\tau_\star \sim 4 \times 10^{-5} M_\odot {\rm yr}^{-1}$ per star.  In constrast, the maintenance of 
a ``immortal'' equilibrium state (with $\lambda_\star \sim 1$, $f_\circlearrowleft=0$,  
\S\ref{mixingsection}, and $M_\star \simeq M_0 =630 M_\odot$, Fig. \ref{fig:diffusion}) 
requires a H-to-He conversion rate 
\begin{equation}
{\dot {\mathcal M}}_{\rm He} = {L_\star \over  \epsilon_{\rm He} c^2} 
= {\lambda_\star M_\star \over \epsilon_{\rm He} \tau_{\rm Sal}} 
\simeq 2 \times 10^{-4} {M_\odot  \over {\rm yr}} \sim 5 {\dot M}_{\rm He}.
\end{equation}
This difference is due to ${\bar M}_{\rm eq} \simeq 0.22 M_0$ so that ${\bar L}_\star \simeq 
{L_{\rm E}}_\star ({\bar M}_{\rm eq}) \simeq 0.22 {L_{\rm E}}_\star (M_0) 
\simeq 0.22 L_\star (M_0)$ during stage 2 of MS evolution (Eq. \ref{eq:averagelum}, \S\ref{sec:starpop}).  
In order for the evolving stars to provide an adequate auxiliary power to maintain 
a state of marginal stability ($Q \sim 1$ with $Q^+ _\star \simeq Q^-$), stars which undergo 
the MS-to-PostMS transition need to be more numerous than the ``immortal'' stars 
(i.e. with a constant $s_\star ({\bar M}_{\rm eq} ) {\bar L}_\star$ determined by $Q^-$
in \S\ref{sec:starpop}).  This compensation
implies that the difference in the total He pollution rate for the disk by the evolving stars
may be indistinguishable from that by the ``immortal'' stars.   Although it is not possible 
to eliminate the possibility of ``immortal'' stars, {\it solely} based on the modest observed value of $Y_{\rm d}$, it does eliminate predominant He pollution as a potential mechanism to enable the MS-to PostMS transition \citep{jermyn2022}.

\begin{figure*}
        \includegraphics[scale=0.48]{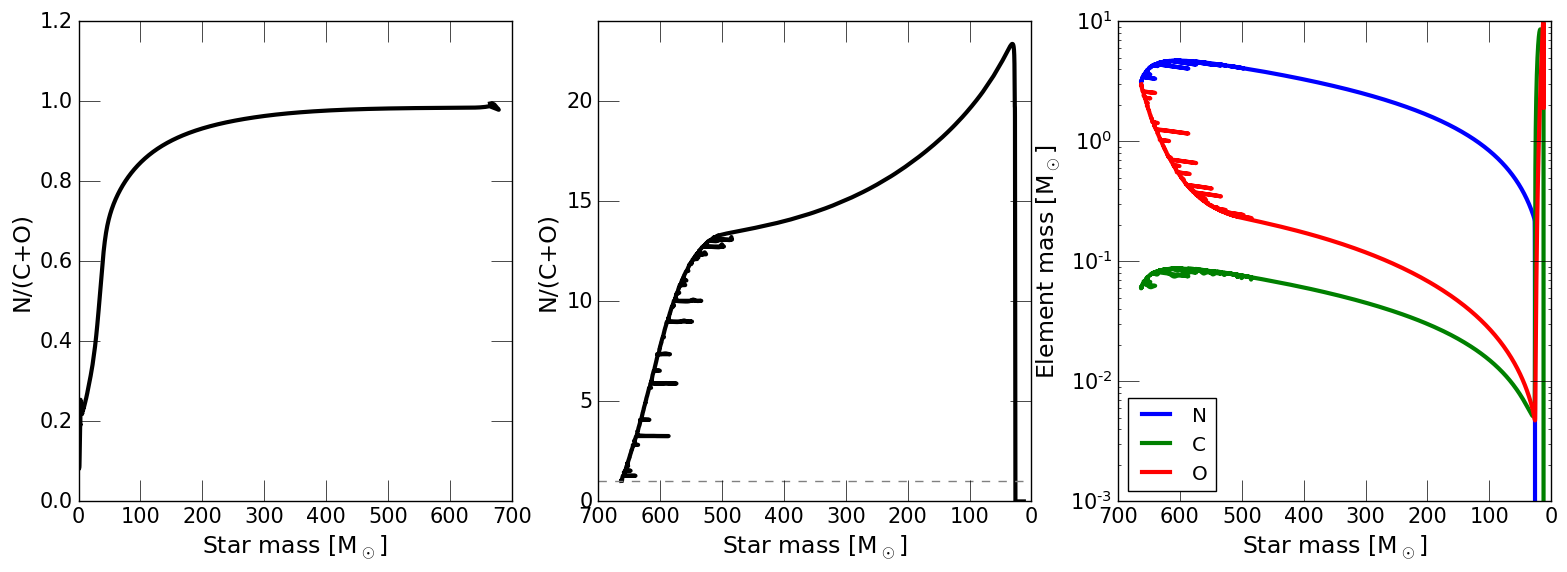}
   \caption{{Left panel:} the evolution of N/(C+O) mass ratio during the initial phase where we grow a 1 M$_\odot$ object embedded in a solar composition disk into a ``immortal'' star. {Central panel:} the evolution of N/(C+O) mass ratio during the subsequent mass loss phase, after local helium retention has been turned on. {Right panel:} the mass evolution curves of C,N,O during the mass loss phase. } 
    \label{fig:cno}
\end{figure*}

\subsection{Pollution rate of $\alpha$ elements.}
\label{sec:productionalpha}
MS stars' luminosity is powered by the CNO process which does not increase the abundance of C+N+O
from that of the disk gas ($\sim 2 \%$).  Nevertheless, it leads to the conversion of C+O to N 
as secondary elements during the initial mass rump up to $M_0$ and subsequent MS stages 1 and 2
mass loss to $M_a$ and $M_2$ respectively (left and middle panels of Fig. \ref{fig:cno}) 
with substantial increase in N/(C+O) to super solar
values ($\gtrsim 20$).  Along with He, 5.1 M$_\odot$ of N and negligible amounts of C and O are returned to
the disk during the MS stages 1 and 2 of evolving stars (left panel of Fig. \ref{fig:yield} and 
\S\ref{sec:secondgen}).  The N/(C+O) ratio may increase further for the immortal stars if
they re-accrete gas already laden with secondary byproducts (i.e. super solar N/(C+O)).

The main difference between the ``immortal'' and the evolving stars is in the production of
$\alpha$ elements and Fe during the PostMS evolution of the latter.  
As they make a transition between stages 2 and 3, the evolving stars are 
primarily composed of He.  During PostMS stage 3, He is converted into $\alpha$ at a rate ${\dot M}_\alpha 
\simeq L_\star / \epsilon_{\rm He} c^2 = \lambda_\star M_\star / \epsilon_\alpha \tau_{\rm Sal}$.
On the timescale $\Delta t_{\rm PostMS} \sim M_\star/{\dot M}_\alpha \simeq \epsilon_\alpha \tau_{\rm Sal} 
\sim 0.5$ Myr, all the He contents would be exhausted (\S\ref{sec:equimassmu})
in agreement with MESA model (Fig. \ref{fig:main}).  Although this time scale
is an order of magnitude longer than the super-Eddington $M_\star/{\dot M}_{\rm wind}$
(with $\lambda \gtrsim 1$ and $A_\lambda \sim 1$ in Eq. \ref{eq:windmod0}), the duration of stage 3 
is prolonged by the quasi accretion-wind equilibrium (with $\lambda_\star \lesssim 1$, Fig. \ref{fig:main}).
Despite the re-accretion of the disk gas, the presence of an outer radiative region (Fig. \ref{fig:main2a})
prevent the replenishment of fresh fuel to the He-burning core.  The net reduction of the PostMS stars
$M_{\rm eq}$ is due to the ramp up of $\alpha$, Fe, and $\mu_\star$, analogous to the quasi-equilibrium
mass loss during the MS evolution (\S\ref{sec:mstarmu}).

{As calculated in Appendix \ref{sec:equimassmu}, during PostMS stage 3,} $A_\mu \simeq -1.5$, $X_\star=0$, $\mu_\star \simeq 1.3 \rightarrow 2$,
$A_{\rm z} \equiv {\partial {\rm ln} M_{\rm eq} 
/ \partial {\rm ln} Z_\star }
=A_\mu \partial {\rm ln} \mu_\star/\partial {\rm ln} Z_\star \simeq 
A_\mu Z_\star/(3-Z_\star) \simeq -0.01 \rightarrow -0.75$, the equilibrium mass 
\begin{equation} 
M_{\rm eq} \simeq M_2 {\rm exp} A_{\rm z} \simeq M_2 {\rm exp} (A_\mu Z_\star/(3-Z_\star)) \simeq  28 \rightarrow 14
\label{eq:mstage3}
\end{equation}
as $Y_\star \simeq 1 \rightarrow 0$ and $Z_\star \simeq 0.02 \rightarrow 1$.  
The rate of $\alpha$ and Fe elements return to the disk is ${\dot M}_{\rm Z}=- Z_\star {\dot M}_{\rm eq}$ 
where ${\dot M}_{\rm eq} = A_{\rm z} M_{\rm eq} {\dot Z}_\star / Z_\star$ is the changing 
rate of the equilibrium mass $M_{\rm eq} (Z_\star)$ 
(Eq. \ref{eq:mstage3}).  Neglecting variations in $A_{\rm \mu}$, the total $Z_\star$ return to the disk is
\begin{equation}
    \Delta M_{\rm Z} = -\int_{0.02} ^1  {A_{\mu} Z_\star M_2 \over (3 - Z_\star)} {\rm exp}
    {A_z Z_\star \over (3-Z_\star)}   d Z_\star 
    \sim  0.2 M_2 \sim 6 M_\odot
    \label{eq:zreturn}
\end{equation}
which is comparable to the heavy element return from the MESA model (
$\sim$ 4.3 M$_\odot$ of C and 3 M$_\odot$ of O, along with small amounts of 
Ne and virtually no N, left panel of Fig. \ref{fig:yield}).
A fraction of the stellar mass $M_2-\Delta M_{\rm Z}-M_3 \simeq 9 M_\odot$ is released as He.  
Most of the $\alpha$ ($9.6 M_\odot$ O, $0.6 M_\odot$ Ne, $1.6 M_\odot$ Mg, $0.7M_\odot$ Si) 
and Fe are contained in the pre-collapsing core (with 
a total mass $M_3\simeq 13 M_\odot$) at the end of stage 3 (right panel of Fig. 
\ref{fig:yield}).  A fraction of $M_3$ is released as $\alpha$ and Fe ejecta 
(with sub-solar [Fe/$\alpha$]) from the subsequent 
SNII and the remainder is contained in the collapsed rBHs \citep{sukhbold2016}. {Note that the mass of the rBH is assumed to be the mass of the star at the end of the MESA simulation, when core collapse is starting as seen in Fig. 2 (central panel). This mass is of course not necessarily the final mass of the remnant, so it should be taken as an upper limit.} The total mass of
extra $\alpha$ return ($\sim \Delta M_{\rm Z}+M_3 \simeq 20 M_\odot$, Fig. \ref{fig:yield}) is $\sim 3\%$ that
of H+He, i.e. $\Delta Z \sim 1.5 Z_{0}$, leading to the $\alpha$ enrichment of the disk gas.

The ratio $\Delta Y/\Delta Z \sim 9$ which is substantially above that ($\sim 1.8$) inferred from
galactic chemical evolution\citep{carigi2008, peimbert2016, valerdi2021}.  However, the magnitude 
$\Delta M_{\rm He}$ (and therefore $\Delta Y$) of individual evolving star would be reduced if the initial 
equilibrium mass ($M_0$) was limited by gap formation with a mass $M_{\rm gap} < M_{\rm eq} (X_0, Y_0, Z_0)$
(\S\ref{sec:modroche}). Moreover, the total $\alpha$ and Fe production rate is proportional to ${\dot N}_\star 
\Delta M_{\rm Z} \simeq N_\star \Delta M_{\rm Z} / \tau_\star$ and the population ($s_\star$ and $N_\star$) 
of coexisting evolving stars have much larger (by a factor $\sim M_0/{\bar M}_{\rm eq} \sim 4.4$ in 
the fiducial model) than those of the ``immortal 
stars'' population (\S\ref{sec:starpop} \& \ref{sec:diskcomp}).  But the H-to-He conversion rate is 
regulated by the auxiliary power criterion which does not depend on the stars' $M_\star$ and $N_\star$
(\S\ref{sec:diskcomp}).
Consequently $\Delta Y/\Delta Z$ is much reduced from that inferred from the nominal model,
and more in line with the observed values \citep{2023MNRAS.tmp.2527H}.

\subsection{Multiple generations of evolving stars}
\label{sec:secondgen}
The pre-collapse cores of evolving stars contain the byproducts 
of triple-$\alpha$ reaction during their PostMS stage 3 which 
are mostly O with negligible amount of N.
However, the byproducts of CNO burning during the MS evolution of
subsequent generation of stars efficiently converts O+C into N
as secondary products.  This conversion is evident in the ramp 
up to $m_{\rm eq} = 630$.  Although the disk gas is assumed to have a solar
composition (with N/(C+O) $\lesssim 0.08$), CNO burning substantially
ramp up the stellar N abundance (with N/(C+O) $\sim 20$ in $M_0$, 
Fig. \ref{fig:yield}).  During MS stages 1 and 2, the N return
is an order of magnitude larger than the O+C return to the disk.
With additional $\alpha$ byproducts during the PostMS stage 3, 
the total N/(C+O) in the returned gas is $\sim$ 0.7, which 
is an order of magnitude larger than its solar value. 
This secondary production process is in good agreement
with that observed \citep{edmunds1978, barbuy1983, tomkin1984, thuan1995, henry2000}
and expected from the galactic chemical evolution
\citep{talbot1974, arnett1996, meynet2002}.

\section{Summary and discussions}
\label{sec:summary}
In the high-density environment of AGN disks, embedded stars readily accrete gas and gain mass.  The growth is 
halted either by the formation of gap through tidal truncation or the onset of mass loss through intense wind 
when their luminosity due to nuclear burning approaches the Eddington limit.  The latter stellar metabolism 
generally leads to a high-equilibrium mass.  If the re-accreted gas is well mixed with the nuclear burning
core, these ``immortal'' stars remain on the MS and do not contribute to the super solar $\alpha$
and Fe abundance commonly observed in AGNs. 

In an attempt to resolve this discrepancy between theory and observation, 
we investigate the effect of re-accretion of material which is pre-ejected by the star.
The main outstanding issues are whether embedded stars can be preserved as ``immortal stars'' 
and sufficient conditions for them to undergo transition from MS to PostMS evolution. 

We carry out numerical simulations with the MESA code.  Adopting the same prescriptions
on accretion rate, wind flux, boundary conditions, and internal structure as \citep{cantiello1}, 
we reproduce previous models for ``immortal stars'' in the dense inner region and for {\it evolving} 
stars in the outer tenuous regions (beyond a few AU) of AGN disks.  We then modify our MESA model
with: 1) the adaptation of the solar composition rather than prestine composition for the AGN disks,
2) the suppression of mixing in the radiative region, 3) the introduction of a retention factor, and 
4) the consideration of a upper-limit for the stellar mass due to gap formation.
We found effects 2, 3, and 4) can lead to the transition from ``immortal'' to {\it evolving} stars.

High retention efficiency recyles most He byproducts on the MS and it is equivalent to 
a ``closed-box'' model.  After creating an ``immortal'' star with MESA and then turning 
on 95\% retention efficiency, we find that this set of initial and boundary conditions 
causes the stars to lose mass on a chemical evolution timescale of few Myr.  This net mass
loss is due to the gradual He accumulation and molecular weight increase which change the
$L_\star-M_\star$ relationship and the value of $M_{\rm eq}$ when the Eddington limit is
approached. The star runs out of H as its $M_{\rm eq}$ is reduced to $\sim 30M_\odot$
and it transitions to PostMS evolution with the onset of triple-$\alpha$ reaction.
Subsequently, nuclear synthesis generates $\alpha$ elements through $\alpha$ chain reaction.
This model shows that with high retention efficiency ``immortal'' stars are 
impermanent and have a short MS lifetime of few Myr.

In relatively-thin AGN disks \citep{starkey2022}, the embedded stars' tidal torque leads to gap formation 
and their mass reaches an upper-growth-limit.  Provided this mass is less than $\lesssim 600-700 M_\odot$,
the evolving stars have a radiative zones in their envelopes which can prevent compositional mixing
between the nuclear burning cores and the surface region where gas is being exchanged between the stars 
and the disk.  This insulation also enables the accumulation of He ashes, 
quenches H replenishment, and lead to the MS-to-PostMS transition on the time scale of a few Myr.

During the PostMS stage, the evolving stars continue to loss mass as their $\mu_\star$ increases further 
until silicon is ignited when their mass is reduced to $\sim 13-15 M_\odot$.  Due to technical limitations, 
we terminate our calculation.  By the end of their PostMS track, individual {\it evolving} stars would have 
ejected hundreds of M$_\odot$ of He, several M$_\odot$ of $\alpha$ elements, and a fraction of $M_\odot$ of Fe 
back into the AGN disks, significantly affecting their chemical composition.  When the polluted disk gas
is being recycled, reprocessed, and returned by the next generation stars, their N/(C+O) ratio is reset by the CNO reaction
on the MS track.  These byproducts are mixed with the disk gas and together they diffuse toward the SMBH at the center
of AGN disks.  In combination, these processes determine the abundance distribution of AGN disks.

\subsection{Uncertainties of the numerical model}
\label{sec:numsimp}

In our numerical model multiple simplifications were used to enhance numerical stability. 
Here we discuss the effects of relaxing these simplifications. In Fig. \ref{fig:main} we 
show curves (in orange) where we removed the \textit{tanh} smoothing terms from the accretion 
and mass loss equations, and replace them with a manual condition where $\dot{M}_{\rm Bondi}$ 
is 0 for $L_\star > {L_{\rm E}}_{\star}$, and $\dot{M}_{\rm wind}$ is 0 for $L_\star 
< {L_{\rm E}}_{\star}$. We find that this approach, coupled with a manually-set 
maximum limit on the timestep, is enough to keep the numerical calculations 
stable. The qualitative behavior of the star is unchanged, where it loses mass 
due to helium buildup till it reaches PostMS. Comparing the mass loss curves for the 
nominal and ``numerically unsmoothed'' models,  we find that they reach the same 
final mass around 13 M$_\odot$, although the ``numerically unsmoothed'' model 
reaches it $\sim$ 1.5 Myr earlier. The Eddington factor for the  unsmoothed model 
oscillates more closely around unity as can be seen in the middle panel of Fig. 
\ref{fig:main}, since the star is not allowed to lose mass for $L_\star/{L_{\rm E}}_{\star}$ 
< 1, in contrast with our nominal model. In this case, $L_\star/{L_{\rm E}}_{\star}$ 
keeps oscillating between two states where it gains mass during a step increasing 
its $L_\star /{L_{\rm E}}_\star$ to unity, then loses mass during the next step 
decreasing its $L_\star/{L_{\rm E}}_\star$ to below unity.
{We emphasize that, in reality, no star can exceed the surface Eddington limit. Stars at the Eddington luminosity are not in hydrostatic equilibrium. The reasons our models do briefly exceed ${L_{\rm E}}_\star$, are almost entirely numerical as explained above and not due to the physical assumptions of the model or the treatment of ${L_{\rm E}}_\star$.  Fundamentally, any similar 1D numerical model will suffer from transient Super-Eddington luminosities unless the timestep used is extremely small (hindering long term model evolution). In a hypothetical global fully hydrodynamical 3D simulation, the situation is partly rectified by the 3D flow where the star, at each timestep, can simultaneously accrete mass at the cool equator, and loose mass at the hotter poles. However, even this model is still dependent on timestepping that might lead to transient Super-Eddington luminosities.}

\subsection{Effects of suppressed winds and the $\mu_\star$ gradient}
Here we start by investigating the role the wind efficiency factor $A_\lambda$ 
in Equations (\ref{eq:masslosse}) and (\ref{eq:windmod0}). So far we assumed 
$A_\lambda \sim 1$, implying maximally-efficient winds. In reality, effects 
such as the ``porosity'' of the star's atmosphere and photon tiring can suppress 
the winds efficiency \citep{shaviv}. Manually decreasing $A_\lambda$ 
by a factor 10 in our nominal simulation however does not lead to qualitative 
changes. We find that winds efficiency need to be suppressed by a factor 30-100 
for the star to evolve PostMS. A suppression of this magnitude is unlikely, and thus 
reduced winds efficiency probably does not play a central role in evolving the 
stars to the PostMS phase. However, it is possible that moderately suppressed winds, coupled 
with some other effects such as local wind-retention discussed above, might be 
enough to force the star to evolve through the MS to PostMS phase. 
Finally, we also investigated the effects of using the Ledoux convection criteria 
instead of the default Schwarzschild criteria and allowing for semi-convection, 
and found that this does not affect our nominal model either. 

\subsection{Retention efficiency, mixing barrier, mass growth limit}
 The analyses and simulations presented here highlight two sufficient criteria for MS-to-PostMS 
transition: a) a high retention efficiency or b) a mixing barrier.  We postulate the possibility
that in the limit of $\lambda_\star \sim 1$, there is a region between the Bondi and Roche radius
where the momentum and mass fluxes of the outgoing-wind and incoming-accretion flux are well mixed 
and subsonic.  High retention efficiency is based on the assumed stability of this two-flow pattern.
This hypothesis needs follow-up verification and examination with some 3D radiative hydrodynamic
simulations.

Under our assumption that there is no compositional mixing in the radiative zone.  This structure provides
a buffer between the mass-exchanging outer layers and the nuclear-burning core of the stars.
In contrast, \cite{cantiello1} suggest the possibility of extra mixing due to rotational 
instability.  Rapid spin (a necessary condition for rotational instability (and hence 
compositional mixing throughout the radiative zone) is commonly observed in stand-alone 
massive stars.  Whether similar conditions may be common for the massive stars which 
grow through the accretion of turbulent disk gas is another area of natural follow-up studies
with 3D hydrodynamic simulations.

We introduce a scenario that gas accretion onto embedded stars is quenched with an upper limit on their
masses $M_{\rm gap}$.  For illustration purposes, we choose $M_{\rm gap}=300 M_\odot$ (Fig. \ref{fig:gap}).
Provided this mass is sufficiently modest to preserve a significant radiative
zone, we can bypass the stringent retention criterion and ensure MS-to-PostMS transition for stars
embedded over a large radial range in AGN disks.  
The gap formation criterion in \S\ref{sec:modroche} is derived and simulated for isolated low-mass
companions (such as planets) embedded in their natal disks on nearly circular
orbits \citep{lin1993, nelson2012}.  Tidal interaction between embedded stars and AGN disks endures
perturbation from neighboring coexisting stars.  The effectiveness of mass-growth limit
needs to be further studies and simulated.





\section*{Acknowledgements}
We thank Andrew Cumming for suggestions, guidance, and major contributions throughout this investigation.
We thank Jiamu Huang, Greg Shields, Stan Woosley, Yixian Chen, and Jamie Law-Smith, and Chris Fryer for 
useful conversation. We thank the anonymous referee for their insightful comments that significantly improved the manuscript. This work is supported by Tamkeen under the NYU Abu Dhabi Research Institute grant CAP$^3$.

\section*{Data availability}
The data underlying this article (numerical simulations output files) will be shared 
on reasonable request to the corresponding author.

\appendix

\section{Bondi accretion rate near the Eddington limit.}
\label{sec:modbondi}
We first consider the conventional analysis for Bondi accretion in the limit
that the stars' luminosity is near their Eddington limit.  
The influence of radiation pressure from stellar luminosity on the 
accretion rate and wind flux are separately approximated by 
${\dot M}_{\rm Bondi, \Gamma}$ (Eq. \ref{eq:mdotbondigamma}) 
and ${\dot M}_{\rm wind, \Gamma}$ (Eq. \ref{eq:masslosse}),
in the absence of each other.  Under the assumption that these 
two opposing flows are fully mixed and combined into a single
spherically symmetric radial ($r$) flow, its radial velocity $U_\star$ 
relative to the star (positive/negative values correspond 
to outflow/inflow respectively) is determined by the momentum equation 
\begin{equation}
    {\partial U_\star \over \partial t} + U_\star {\partial U_\star \over \partial r} = 
    -{1 \over \rho} {\partial P_{\rm gas} \over \partial r} - (1 - \lambda_\star)
    {G M_\star \over r^2}
    \label{eq:momentum}
\end{equation}
where the radiation pressure term is incorporated by the $\lambda_\star GM_\star/r^2$ term.  
The magnitude of $\lambda_\star \equiv L_\star/ {L_{\rm E}}_\star$ is an increasing 
function of $m_\star$.  In a steady state (when the mass flux ${\dot M}_{\rm Bondi} =4 
\pi \rho U_\star r^2$ is independent of $r$, Eq \ref{eq:bondiaccrate}), the sonic point 
(where $U_\star=c_{\rm s}$) occurs at the Bondi radius 
\begin{equation}
R_{\rm B} = (1 - \lambda_\star) G M_\star / 2 c_s^2.
\label{eq:bondilambda}
\end{equation}
Stellar wind is launched from MS stars' 
${\rm R}_\star \ll R_{\rm B}$ when $\lambda_\star < < 1$.  But, with $\lambda_\star \rightarrow 1$, 
$R_{\rm B} \rightarrow{\rm R}_\star$, the entire accretion flow becomes subsonic (with $U_\star 
\lesssim c_{\rm s}$), and the effective accretion rate is greatly quenched
\begin{equation}
{\dot M}_{\rm Bondi} \simeq 4 \pi \rho_{\rm c} c_s R_{\rm B}^2
\simeq 2.5 \times 10^{20} (1 - \lambda_\star)^2 
{\alpha_\nu m_\star^2 \over f_\bullet r_{\rm pc} ^{3}} \ {\rm g \over s}
\label{eq:bondiacc}
\end{equation}
where $c_{\rm s}$ and $\rho_{\rm c}$ are obtained from Equations (\ref{eq:hscale}) and (\ref{diskdeneq}).
This derivation provides the basis for the prescription for Equation (\ref{eq:mdotbondigamma})

\section{Accretion-wind equilibrium}
\label{sec:gap}
A simplifies prescription for the wind 
\begin{equation}
{\dot M}_{\rm wind} \simeq {A_{\lambda} L_\star R_\star \over G M_\star} \simeq \lambda_\star A_{\lambda} {m_\star^{0.6} 
M_\odot \over \tau_{\rm Sal}} {R_\odot \over r_\bullet} \simeq 7 \times 10^{22} \lambda_\star A_\lambda m_\star ^{0.6} 
{\rm g \over s}
\label{eq:windmod0}
\end{equation}
where $R_\star \simeq m_\star ^{0.6} R_\odot$ and $R_\odot$ are the radius of star and the Sun,
$r_\bullet = G M_\odot/c^2$ and $M_\odot$ are the gravitational radius and mass of the Sun,
and $\tau_{\rm Sal} (= {L_{\rm E}}_\odot/ M_\odot c^2= 4.5 \times 10^8$ yr) 
is the Salpeter time scale. In comparison with Equation (\ref{eq:masslosse}), the factor $A_{\lambda} =
[ 1+\tanh (10\lambda_\star-10)]/2$ which $=0, 0.5,$ and 1 for $\lambda_\star << 1$, 1, and $>> 1$ respectively.

An accretion-wind equilibrium (${\dot M}_{\rm Bondi} = {\dot M}_{\rm wind}$) is attained with 
\begin{equation}
    {(1-\lambda_\star)^2 \over \lambda_\star A_{\lambda}} \simeq {2.8 \times 10^2 f_\bullet r_{\rm pc}^3 \over \alpha_\nu m_\star ^{1.4}}.
    \label{eq:oneminuslam}
\end{equation}
For $m_\star \sim 600-1000$ at $r_{\rm pc} \sim 1$, $\lambda_\star \sim 0.9-1$ and these results are consistent with the asymptotic 
equilibrium mass $M_{\rm eq}$ and Eddington factors computed with the MESA models without recycling $f_\circlearrowleft =0$ (Fig. \ref{fig:diffusion}).
In this equilibrium (with $M_\star=M_{\rm eq}$), the Bondi accretion time scale $\tau_{\rm Bondi} \equiv {M_\star / {\dot M}_{\rm Bondi}}$
(Eqs \ref{eq:bondiacc} and \ref{eq:windmod0}) becomes
\begin{equation}
\tau_{\rm Bondi} \simeq 
{0.25 f_\bullet r_{\rm pc} ^{3} {\rm Myr} \over 
(1 - \lambda_\star)^2 \alpha_\nu m_\star} = {M_\star \over {\dot M}_{\rm wind}} 
= {\tau_{\rm Sal} m_\star^{0.4} \over A_\lambda \lambda_\star }
{r_\bullet \over R_\odot}.
\label{eq:taubondisal}
\end{equation}

\section{Roche-lobe flow modified by radiation pressure}
\label{sec:modroche}
Analogous to the emergence of proto-gas-giant planets, the embedded stars also tidally interact
with the disk near their Roche/Hills radius 
\begin{equation}
R_{\rm H} = (1-\lambda_\star)^{1/3} 
(M_\star/3 M_\bullet)^{1/3} R.
\label{eq:rochelambda}
\end{equation}
This equation takes into account 
the reduction of the star's gravity by
the radiation pressure imposed by its luminosity.
At relative small radii, it is possible for 
embedded stars' $M_\star$ towards an $M_{\rm eq} \gtrsim 3^{1/3} h^3 
M_\bullet /(1-\lambda_\star)$ with $\lambda_\star < 1$. But their growth would be quenched
\citep{dangelo2003, dobbsdixon2007, bodenheimer2013,  tanigawa2016, rosenthal2020, lichenlin2021} by the 
formation of a gap in the proximity of their orbit 
\citep{lin1993} when their mass $M_\star \rightarrow M_{\rm gap} \simeq h^3 M_\bullet$. 
Substitute $h$ (Eq. \ref{eq:hscale}), gaps form when their 
normalized $m_\star$ exceeds
\begin{equation}
    m_{\rm gap} = {M_{\rm gap} \over M_\odot} 
    \simeq {10^3 f_\bullet m_8^{1/2} r_{\rm pc}^{3/2} \over  \alpha_\nu (1-\lambda_\star)}.
\label{eq:gap}
\end{equation}
This gap-opening process also reduces the local $\Sigma_{\rm g}$, $\rho_{\rm c}$  
\citep{duffell2013, kanagawa2015, chen2020} and limits stars' asymptotic $R_{\rm B} \sim R_{\rm H} \sim H$
and impede their accretion with $M_\star \sim M_{\rm gap} < M_{\rm eq}$ and $\lambda_\star$ $< < 1$.  

At relatively large $R$, embedded stars continue to gain mass until $m_\star 
\rightarrow m_{\rm eq}$ with $\lambda_\star \simeq 1$.  In this limit, 
$m_{\rm gap} \gtrsim m_{\rm eq}$, gap formation does not occur, and in the absence
of radiative feedback, 
$G M_{\rm eq}/c_s^2 < (M_{\rm eq}/3 M_\bullet)^{1/3} R < H$.   
Although $R_{\rm H}$ also retreats as the stars's luminosity reaches
its Eddington limit (i.e. $\lambda_\star \rightarrow 1$), it does not decrease 
as steeply as $R_{\rm B}$ (Eq. \ref{eq:bondilambda} and \ref{eq:rochelambda}). 
When ${\dot M}_{\rm Bondi}$ is limited by 
${\dot M}_{\rm wind}$ with $\lambda_\star \sim 1$ and $R_{\rm B}$ reduced 
to $\sim R_\star$ while $R_{\rm H}$ remains $> R_{\rm B}$. 
We adopt the approximation $f_\circlearrowleft \simeq 1$ (\S\ref{sec:reaccrete}) under the assumption that these 
embedded stars would recycle the He polluted gas in the surrounding, nearly-static, 
pressure-supported envelope between $R_{\rm B}$ and $R_{\rm H}$ that is 
tidally bound to them.  The fully mixed wind-accretion assumption needs to 
be verified by future investigations.

\section{Critical retention efficiency}
\label{sec:analyticrecycle}
An accretion-wind equilibrium is maintained with an influx of fresh disk 
gas at a rate $(1-f_\circlearrowleft) {\dot M}_{\rm Bondi}$. The  
rate of change in the total hydrogen mass due to accretion and wind is
\begin{equation}
    {\dot M}_{\rm H, \circlearrowleft}  \simeq 
    (1- f_\circlearrowleft) (X_{\rm d} - {\rm X}_\star) M_\star/\tau_{\rm Bondi}. 
\label{eq:mxdot0}
\end{equation}
 
To power $L_\star$ on the MS, H is also converted to He with an efficiency 
$\epsilon_{\rm He} (\simeq 0.007)$. The net H mass change rate is
\begin{equation}
    {\dot {M}}_{\rm H, net} = 
    {\dot M}_{\rm H, \circlearrowleft}  - {L_{\star} \over \epsilon_{\rm He} c^2}
    ={\dot M}_{\rm H, \circlearrowleft}  - {\lambda_\star M_\star \over \epsilon_{\rm He} \tau_{\rm Sal}}.
\label{eq:totalxdot}
\end{equation}
In addition, changes in composition ($X_\star$ and $\mu_\star$) modify the equilibrium mass 
(\S\ref{sec:mstarmu} and \S\ref{sec:equimassmu})
\begin{equation}
{{\dot M}_\star \over M_\star} = {{\dot M}_{\rm eq} \over M_{\rm eq}} = {A_{\rm x} {\dot X}_\star 
\over X_\star}, \ \ \ \ \ \ A_\mu \equiv {d {\rm ln} M_{\rm eq} \over \partial {\rm ln} \mu_\star} 
\nonumber
\end{equation}
\begin{equation}
M_{\rm eq} \propto X_\star ^{A_{\rm x}}, \ \ \ \ {\rm and} \ \ \ \ 
A_{\rm x} \equiv {d {\rm ln} M_{\rm eq} \over \partial {\rm ln} X_\star}
=A_\mu {{\partial {\rm ln} \mu_\star} 
\over \partial {\rm ln} X_\star},
\label{eq:dlnmdlnmu}
\end{equation}
is a positive factor of order unity (Fig. \ref{fig:helium} and \S\ref{sec:equimassmu}). The net compositional change 
\begin{equation}
{\dot X}_\star = {{\dot M}_{\rm H, net} \over M_\star} - {X_\star {\dot M}_\star \over M_\star}
={{\dot M}_{\rm H, \circlearrowleft} \over M_\star}  - {\lambda_\star \over \epsilon_{\rm He} \tau_{\rm Sal}}
- {X_\star {\dot M}_{\rm eq} \over M_{\rm eq}}.
\label{eq:dotxstar} 
\end{equation}
From Equations (\ref{eq:mxdot0}), (\ref{eq:dlnmdlnmu}), and (\ref{eq:dotxstar}), it follows
\begin{equation}
     {\dot X}_\star
    = { 1\over (1 + A_{\rm x})} \left( {(1- f_\circlearrowleft ) (X_{\rm d} - {\rm X}_\star) \over \tau_{\rm Bondi}}
    - {\lambda_\star \over \epsilon_{\rm He} \tau_{\rm Sal}} \right).
    \label{eq:xdotstar}
\end{equation}
Inadequate replenishment of H (first term inside the bracket) to compensate the convertion of H to He (second term
inside the bracket) would lead to a decline in $X_\star$ (i.e. ${\dot X}_\star <0$) and stars' equilibrium mass
(i.e. ${\dot M}_\star <0$).

With a substitution for $\lambda_\star \tau_{\rm Bondi}/\tau_{\rm Sal}$ (Eq. \ref{eq:taubondisal}),  
${\dot X}_\star <0$ ($X_\star$ declines well below $X_{\rm d}=0.7$ over time) provided
\begin{equation}
    1-f_\circlearrowleft \lesssim {\lambda_\star \tau_{\rm Bondi}/\tau_{\rm Sal} \over \epsilon_{\rm He}  (X_{\rm d} - X_\star)}
    \simeq  {m_\star ^{0.4} (r_\bullet/R_\odot) \over \epsilon_{\rm He} (X_{\rm d} - X_\star)   A_\lambda}
    \simeq {4.3 \times 10^{-4} m_\star ^{0.4} \over   A_\lambda}.
\label{eq:xdeclinecon}
\end{equation}
With $A_\lambda \simeq 0.5$ (for $\lambda_\star \simeq 1$) and $m_\star \sim 600$ (from MESA models in 
\S\ref{mixingsection}), we find the condition for total H depletion is $f_\circlearrowleft \simeq 0.99$.

If we set $f_\circlearrowleft=0.95$  as shown \S\ref{sec:reaccrete}, a minimum $X_\star \sim 0.5$ would be 
reachable before ${\dot X}_\star$ vanishes (Eq. \ref{eq:xdotstar}).  Since ${\dot Y}_\star = - {\dot X}_\star$, 
the magnitude of $\mu_\star$ increases and the equilibrium mass $M_{\rm eq}$ decreases $\lesssim 485$
with the He enhancement (Eq. \ref{eq:oneminuslam}).  Below this mass, further decline 
in the $X_\star$  leads to a rapidly expanding radiative region in 
the outer envelope of the star (Fig. \ref{fig:main2a} \& \S\ref{sec:mstarmu}).  This zone segregates the replenished gas 
and the stellar interior which is equivalent to resetting insulation with $f_\circlearrowleft=1$ for the
nuclear burning core.  This internal structure modification enables the star to undergo MS-to-PostMS
transition.  But with $f_\circlearrowleft \lesssim 0.9$, the $X_\star$ decline is quenched with $X_\star$ only slightly
less than $X_{\rm d}$ while $m_\star \gtrsim 485$ with negligible radiative barrier to prevent the mixing
of the accreted material and the burning core.  In this case, stars may preserve their longevity. 

\section{Equilibrium-mass and molecular-weight evolution}
\label{sec:equimassmu}
In \S\ref{sec:mstarmu}, we attribute the evolution of equilibrium mass obtained from
the MESA model to increases in the molecular weight $\mu$.  For an analytic approximation, 
we start with the definition of the radiative pressure gradient:
\begin{equation}
    {dP_{\rm rad} \over dr} = -\frac{\kappa_\gamma \rho_\star}{4\pi c}\frac{L(r)}{r^2}
\end{equation}
which taken with $dP/dr = -GM\rho_\star/r^2$ gives:

\begin{equation}
{dP_{\rm rad} \over dP} = \frac{\kappa_\gamma L(r)}{4\pi GMc} \simeq \frac{L_\star}{{L_{\rm E}}_\star}.
\end{equation}
With $P=P_{\rm rad}+P_{\rm gas}$ and $\beta = {P_{\rm gas}}/{P} = 1-{P_{\rm rad}}/{P}$,  
\begin{equation}
\label{1mbeta}
    {dP_{rad} \over dP} = 1-\beta \simeq \frac{L(r)}{{L_{\rm E}}_\star }= \lambda_\star
\end{equation}
where $\kappa_\gamma$ is the opacity inside the star.
For an ideal gas $P_{\rm gas} = {\rm R_g} \rho_\star T_\star/\mu_\star$ where $R_{\rm g}$ is the gas constant and
the molecular weight for fully ionized gas $\mu_\star \simeq 4/(3+5X_\star-Z_\star)=4/(8-5Y_\star-6Z_\star)$ reduces to
\begin{equation} 
\mu_\star \simeq {4 \over 8-5 Y_\star}, \ \ \ \ {\partial {\rm ln} \mu_\star \over \partial {\rm ln} Y_\star} 
\simeq{5 Y_\star \over 8-5 Y_\star}, 
\ \ \ \ {d {\rm ln}\mu_\star \over d {\rm ln} X_\star} \simeq {-5 X_\star \over 3 + 5 X_\star}
\label{eq:mux1z0}
\end{equation}
during the initial MS evolution when $Z_\star < < 1$. 
During the PostMS with  $X_\star < < 1$,
\begin{equation} 
\mu_\star \simeq {4 \over 3-Z_\star}, \ \ \ \ {\partial {\rm ln} \mu_\star \over \partial {\rm ln} Z_\star} 
\simeq{Z_\star \over 3-Z_\star},
\ \ \ \ {d {\rm ln}\mu_\star \over d {\rm ln} Y_\star} \simeq {- Y_\star \over 2 + Y_\star}, 
\nonumber
\end{equation}
\begin{equation}
M_{\rm eq} \propto Y_\star ^{A_{\rm y}}, \ \ \ \ {\rm and} \ \ \ \
A_{\rm y}  \equiv {\partial {\rm ln} M_{\rm eq} 
\over \partial {\rm ln} Y_\star }
=A_\mu {\partial {\rm ln} \mu_\star 
\over \partial {\rm ln} Y_\star }
\label{eq:mux0z1}
\end{equation}
which is negative $\sim {\mathcal O} (-1)$. 

With radiation pressure $P_{\rm rad}= a T_\star^4/3$). 
and radiation transfer, $F_{\rm rad} =(ac / 3 \kappa_\gamma \rho_\star) d T_\star^4/dr $), \citet{owocki2012} 
adopt the Eddington approximation with scaling relations $\rho_\star (r) \sim 3 M_\star/4 \pi {\rm R}_\star^3$, 
$T_\star \sim (1 - \lambda_\star) \mu_\star G M_\star/ R_{\rm g} {\rm R}_\star$ for a n=3 polytrope, 
$L_\star \sim 4 \pi F_{\rm rad} {\rm R}_\star^2$, and all scale heights 
are comparable to ${\rm R}_\star$ and derive
\begin{equation}
    {\lambda_\star / (1 - \lambda_\star)^4} \propto \mu_\star^4 M_{\star}^2
\label{eq:mudependence0}
\end{equation}
where the $\mu_\star$-dependent opacity $\kappa_\star$ drops out in the above expression.
With the accretion-wind equilibrium (${\dot M}_{\rm Bondi} \simeq 
{\dot M}_{\rm wind}$ and $M_\star = M_{\rm eq}$) being maintained at a nearly constant $\lambda_\star$ (slightly less
than unity), the stellar mass $M_{\rm eq} \propto \mu_\star ^{-2}$ (i.e. $A_\mu \simeq -2$).

There are several stages of stellar evolution (\S\ref{sec:diskcomp}).
The $n=1.5$ polytrope model is a good approximation for stage 1 (MS with $m_\star 
\gtrsim 485$) and stage 3 (PostMS with $m_\star \lesssim 28$) as in the case of 
fully convective stars and cores of red giants\citep{chandrasekhar1939, hansen2004}.  
The equilibrium masses obtained from the MESA model (Fig. \ref{fig:helium}) indicate 
$A_\mu \equiv d {\rm ln} m_{\rm eq}/d {\rm ln} \mu_\star \simeq -1.5$ for $m_\star = m_{\rm eq} \gtrsim 500$ (stage 1) 
and $A_\mu \simeq -2.3$ for $m_\star \lesssim 30$ (stage 3) which are in general agreement
with the above analytic approximation.  They also imply $A_{\rm x} \sim 0.7$ for the 
H-to-He conversion (MS stage 1) and $A_{\rm y} \sim 0.8 \rightarrow 0$ for the He-to-$\alpha$
conversion (PostMS, stage 3) lead to the decline of equilibrium masses (Eq. \ref{eq:dlnmdlnmu}, 
\ref{eq:mux1z0}, \ref{eq:mux0z1}, \ref{eq:mudependence0}, and Fig. \ref{fig:main}).

But for stage 2 (MS with $m_\star \simeq 28-482$), the $n=3$ polytrope model is a better approximation.
For any polytropic star (with $P=K \rho^{(1+n)/n}$), the mass can be written as:
\begin{equation}
\left.M_\star =-\frac{(n+1)^{3 / 2}}{\sqrt{4 \pi}} 
\xi_1^2 \frac{d \theta}{d \xi}\right)_{\xi_1}
\left(\frac{K}{G}\right)^{3 / 2} \rho_{\rm core}^{\frac{3-n}{2 n}}
\end{equation}
where $\rho_{\rm core}$ is the star's core density and
$\xi_1^2 {d \theta}/{d \xi} )_{\xi_1}$ is a constant set by the polytropic index $n$ and 
\begin{equation}
K=\left[\frac{3\left(N_A k_\beta\right)^4(1-\beta)}{a(\mu_\star \beta)^4}\right]^{1 / 3}
\end{equation}
is the adiabatic constant with the Avogadro and Boltzmann constants $N_A$ and $k_\beta$, and the molecular weight for a fully ionized gas is $\mu_\star \simeq 4/(3+5X-Z)$. 
Taking these two equations for n=1.5 polytrope to describe a fully convective star 
with $\rho_{\rm core} \propto M_\star/ R_\star^3$, we find
\begin{equation}
    1-\beta \propto M_\star ^{0.25} R_\star ^{0.75} \mu_\star^4 \beta^4.
    \label{eq:mubeta}
\end{equation}
Equation \ref{1mbeta} and \ref{eq:mubeta} lead to
$L_\star \propto M_\star ^{1.25} R_\star ^{0.75} \mu_\star ^4 \beta^4$. 
Assuming $R_\star \propto M_\star ^{0.6}$ as in our case, one finally gets:
\begin{equation}
    \lambda_\star / (1-\lambda_\star)^4 \propto M_\star ^{0.7} \mu_\star ^4  
\label{eq:lummu}
\end{equation}
such that $\lambda_\star \sim {\mathcal O} (1)$ is maintained 
with $M_{\rm eq} \propto \mu_\star ^{-5.7}$ (i.e. $A_\mu \simeq -5.7$).  This analytic approximation
agrees well with the MESA result for equilibrium mass in the range of $\sim 
30-485$ ($A_\mu \equiv d {\rm ln} m_{\rm eq}/d {\rm ln} \mu_\star \sim -4.7$ during stage 2 in 
Fig. \ref{fig:helium}).  It also implies $A_{\rm x} \simeq 2.1 \rightarrow 0$ 
and $A_{\rm y} \simeq -2.1 \rightarrow -7.8 $ (as $X_\star \simeq 0.5 \rightarrow 0$) 
and H-to-He conversion (MS stage 2) decreases the
equilibrium masses (Eq. \ref{eq:dlnmdlnmu}, \ref{eq:mux1z0}, \ref{eq:lummu},
and Fig. \ref{fig:main}).

Based on the above analytic approximation, we can estimate the time
scale for changing $X_\star$ by an amount $\Delta X_\star$ during MS stage 2 
(as $X_\star \simeq 0.5 \rightarrow 0$, $\mu_\star \simeq 0.72 \rightarrow 1.3$, $A_{\rm x}
\simeq - A_{\rm y} \simeq 2.6 \rightarrow 0$,
and $m_{\rm eq} \simeq 485 \rightarrow 28$) to be $\Delta t_{\rm MS} \sim   
\epsilon_{\rm He} \tau_{\rm Sal}  \sim 3$ Myr (Eq. \ref{eq:xdotstar}).  
This characteristic time scale
agrees well with that obtained with the MESA model.  Minor discrepancies may be due to
the nuclear burning core has only a fraction of the total stellar mass, dependence 
of $A_{\rm mu x}$ on $X_\star$ and $A_{\rm mu y}$ on $Y_\star$, and  
the negligence of the energy needed to power the loss of $M_{\rm eq}$ in the analytic 
approximation (equivalent to a slightly larger $\lambda_\star$ in Eq. \ref{eq:xdotstar}).  
In conclusion, the quasi-equilibrium mass loss time scale is determine
by the chemical-evolution time scale in the core.


\begin{table*}[]
\begin{tabular}{|ll|}
\hline
\multicolumn{1}{|l|}{Variable} & \multicolumn{1}{l|}{Description} \\ \hline
$\dot{M}_{\mathrm{d}}$, $\dot{M}_{\bullet}$, $f_\bullet$, $L_{\rm AGN}$, $\tau_\bullet$ & accretion rates in AGN disks and onto the SMBH, scaling, AGN luminosity, duration\\
$M_\odot$, $R_\odot$, $L_\odot$, $G$, $c$, $z_\gamma$   & Sun's mass, radius, luminosity, gravitational constant, speed of light, and redshift  \\
$\Omega$, $R_{\rm d}$, $R$, $r_{\rm pc}$, $Q$   & AGN disk's angular frequency, size, radius, normalized by pc,
gravitational stability\\ 
$\textit{R}_\bullet$, ${M}_{\bullet}$, $m_8$      & SMBH's gravitational radius, mass in physical units and normalized by $10^8 M_\odot$ \\
$\rho_{\rm core}$, $\rho_c$, $\Sigma_{\rm g}$, $T_{\rm c}$, $T_{\rm e}$     & star's core density, disk's mid-plane and surface density, mid-plane and effective temperature \\
$c_{\rm s}$, $H$, $h$, $\nu$, $\alpha_\nu$      & mid-planet sound speed, thickness, aspect ratio, viscosity, turbulent efficiency \\
$\epsilon_\bullet$, $\epsilon_{\rm He}$, $\epsilon_\alpha$ & energy conversion efficiency due to accretion, hydrogen burning, 
triple-$\alpha$ \& $\alpha$-chain reaction \\
$Q^-$, $Q^+ _\nu$, $Q^+ _\star$, $s_\star$, ${\dot s}_\star$, ${\dot N}_\star$  & disk's cooling flux, viscous, stellar heating 
flux, stars' surface density, formation flux, rate\\
$\kappa_{\rm es}$, $\kappa_{\rm dust}$, $\kappa_\gamma$    & electron scattering, dust, and general opacity \\
$P_\star$, $\rho_\star$, $T_\star$, $F_\star$, $\mu_\star$, ${\dot \mu}_\star$     & stars' internal pressure, density, temperature, radiative flux, molecular weight, changing rate \\
${L_{\rm E}}_{\bullet}$, ${L_{\rm E}}_{\star}$, ${L_{\rm E}}_{\odot}$, $\lambda_\bullet$, 
${\lambda}_\star$, $\tau_{\rm Sal}$       & SMBH's, stars', and Sun's Eddington luminosity and factor, 
Salpeter timescale \\
$V_\star$, $L_\star$, $M_\star$, $m_\star$, $m_{\rm gap}$      & stars' escape speed, luminosity, mass and gap-forming mass normalized by $M_\odot$ \\  
$L_{\rm eq}$, $M_{\rm eq}$, $m_{\rm eq}$,  ${\bar L}_{\rm eq}$, ${\bar M}_{\rm eq}$    & equilibrium luminosity, mass, normalized, 
averaged equilibrium luminosity, mass \\
$\tau_\star$, $\Delta t_{\rm MS}$, $\Delta t_{\rm PostMS}$, $N_\star$  & stars' 
life span, duration of MS H and PostMS He depletion, total population \\
${\rm R}_\star$, $\textit{R}_\star$, ${\rm r}_\star$, $\textit{r}_\star$    & star's physical, gravitational radius, and normalized by solar radius \\
$R_{\rm H}$, ${\dot M}_{\rm Bondi}$, $\tau_{\rm B}$, $R_{\rm B}$, $r_{\rm B}$    & Hill's radius, 
Bondi accretion rate, timescale, radius, and normalized with solar radius    \\ 
$\dot{M}_{\rm wind, \Gamma}$, ${\dot M}_{\rm Bondi, \Gamma}$        
& mass-loss and accretion rates modified by the Eddington limit\\
$\dot{M}_{\rm wind}$, ${\dot M}_{\rm He}$, ${\dot {\mathcal M}}_{\rm He}$, $f_\circlearrowleft$ &
mass loss rate, He release rate by evolving and immortal stars, retention efficiency \\
$A_\lambda$, $\tau_{\rm wind}$, $\rho_{\rm wind}$   
& Wind loss efficiency, time scale, and density at the launch base \\
$P_{\rm gas}$, $P_{\rm rad}$, $U_\star$   & gas and radiation pressure and flow velocity \\
$A_\mu$, $A_{\rm x}$, $A_{\rm y}$, $A_{\rm z}$   & power index of equilibrium-mass distribution with respect to 
$\mu$, $X_\star$, $Y_\star$, and $Z_\star$ \\
$M_0$, $M_1$, $M_2$, $M_3$  & nominal model's initial, onset-of-radiative-zone, PostMS-transition,
pre-collapse masses \\  
${\rm X}_\star$, ${\rm Y}_\star$, ${\rm Z}_\alpha$, ${\rm Z}_{\rm Fe}$, $X_{\rm acc}$   & Stars' H, Helium, 
$\alpha$, and Fe  mass fraction, H mass fraction in accreted gas                 \\ 
$X_0$, $Y_0$, $Y_1$, $Y_2$, $\Delta Y$, $\Delta Z$ & initial H, He fraction, onset-of-radiative-zone, PostMS-transition, change in He, metal fraction\\
$\Delta M_{\rm H}$, $\Delta M_{\rm Z}$, $\Delta M_{\rm He}$
& total mass released to the disk in H, metals, He\\
${\dot{\rm X}}_{\star}$, ${\dot {\rm Y}}_\star$, ${\it X}_{\rm d}$, ${\it Y}_{\rm d}$, ${\mathcal Z}_\alpha$, ${\mathcal Z}_{\rm Fe}$ 
& Net change rates of stars' H and He fraction, disk's H, Helium, $\alpha$, and Fe mass fraction \\
\hline
\end{tabular}
\caption{Table of key variables used in our analytical model. }
\end{table*}

\bibliographystyle{mnras} 
\bibliography{stars.bib}

\begin{thebibliography}{}
\makeatletter
\relax
\def\mn@urlcharsother{\let\do\@makeother \do\$\do\&\do\#\do\^\do\_\do\%\do\~}
\def\mn@doi{\begingroup\mn@urlcharsother \@ifnextchar [ {\mn@doi@}
  {\mn@doi@[]}}
\def\mn@doi@[#1]#2{\def\@tempa{#1}\ifx\@tempa\@empty \href
  {http://dx.doi.org/#2} {doi:#2}\else \href {http://dx.doi.org/#2} {#1}\fi
  \endgroup}
\def\mn@eprint#1#2{\mn@eprint@#1:#2::\@nil}
\def\mn@eprint@arXiv#1{\href {http://arxiv.org/abs/#1} {{\tt arXiv:#1}}}
\def\mn@eprint@dblp#1{\href {http://dblp.uni-trier.de/rec/bibtex/#1.xml}
  {dblp:#1}}
\def\mn@eprint@#1:#2:#3:#4\@nil{\def\@tempa {#1}\def\@tempb {#2}\def\@tempc
  {#3}\ifx \@tempc \@empty \let \@tempc \@tempb \let \@tempb \@tempa \fi \ifx
  \@tempb \@empty \def\@tempb {arXiv}\fi \@ifundefined
  {mn@eprint@\@tempb}{\@tempb:\@tempc}{\expandafter \expandafter \csname
  mn@eprint@\@tempb\endcsname \expandafter{\@tempc}}}

\bibitem[\protect\citeauthoryear{{Arnett}}{{Arnett}}{1996}]{arnett1996}
{Arnett} D.,  1996, {Supernovae and Nucleosynthesis: An Investigation of the
  History of Matter from the Big Bang to the Present}

\bibitem[\protect\citeauthoryear{{Artymowicz}, {Lin}  \&
  {Wampler}}{{Artymowicz} et~al.}{1993}]{artymowicz1993}
{Artymowicz} P.,  {Lin} D.~N.~C.,   {Wampler} E.~J.,  1993, \mn@doi [\apj]
  {10.1086/172690}, \href
  {https://ui.adsabs.harvard.edu/abs/1993ApJ...409..592A} {409, 592}


\bibitem[\protect\citeauthoryear{Huang, Lin, \& Shields}{2023}]{2023MNRAS.tmp.2527H} Huang J., Lin D.~N.~C., Shields G., 2023, MNRAS.tmp. doi:10.1093/mnras/stad2642


\bibitem[\protect\citeauthoryear{{Barbuy}}{{Barbuy}}{1983}]{barbuy1983}
{Barbuy} B.,  1983, \aap, \href
  {https://ui.adsabs.harvard.edu/abs/1983A&A...123....1B} {123, 1}

\bibitem[\protect\citeauthoryear{{Bentz} et~al.,}{{Bentz}
  et~al.}{2013}]{bentz2013}
{Bentz} M.~C.,  et~al., 2013, \mn@doi [\apj] {10.1088/0004-637X/767/2/149},
  \href {https://ui.adsabs.harvard.edu/abs/2013ApJ...767..149B} {767, 149}

\bibitem[\protect\citeauthoryear{{Bodenheimer}, {D'Angelo}, {Lissauer},
  {Fortney}  \& {Saumon}}{{Bodenheimer} et~al.}{2013}]{bodenheimer2013}
{Bodenheimer} P.,  {D'Angelo} G.,  {Lissauer} J.~J.,  {Fortney} J.~J.,
  {Saumon} D.,  2013, \mn@doi [\apj] {10.1088/0004-637X/770/2/120}, \href
  {https://ui.adsabs.harvard.edu/abs/2013ApJ...770..120B} {770, 120}

\bibitem[\protect\citeauthoryear{{Cantiello}, {Jermyn}  \& {Lin}}{{Cantiello}
  et~al.}{2021}]{cantiello1}
{Cantiello} M.,  {Jermyn} A.~S.,   {Lin} D. N.~C.,  2021, \mn@doi [\apj]
  {10.3847/1538-4357/abdf4f}, \href
  {https://ui.adsabs.harvard.edu/abs/2021ApJ...910...94C} {910, 94}

\bibitem[\protect\citeauthoryear{{Carigi} \& {Peimbert}}{{Carigi} \&
  {Peimbert}}{2008}]{carigi2008}
{Carigi} L.,  {Peimbert} M.,  2008, \rmxaa, \href
  {https://ui.adsabs.harvard.edu/abs/2008RMxAA..44..341C} {44, 341}

\bibitem[\protect\citeauthoryear{{Chandrasekhar}}{{Chandrasekhar}}{1939}]{chandrasekhar1939}
{Chandrasekhar} S.,  1939, {An introduction to the study of stellar structure}

\bibitem[\protect\citeauthoryear{{Chatzopoulos} \& {Wheeler}}{{Chatzopoulos} \&
  {Wheeler}}{2012a}]{pair3}
{Chatzopoulos} E.,  {Wheeler} J.~C.,  2012a, \mn@doi [\apj]
  {10.1088/0004-637X/748/1/42}, \href
  {https://ui.adsabs.harvard.edu/abs/2012ApJ...748...42C} {748, 42}

\bibitem[\protect\citeauthoryear{{Chatzopoulos} \& {Wheeler}}{{Chatzopoulos} \&
  {Wheeler}}{2012b}]{pair2}
{Chatzopoulos} E.,  {Wheeler} J.~C.,  2012b, \mn@doi [\apj]
  {10.1088/0004-637X/760/2/154}, \href
  {https://ui.adsabs.harvard.edu/abs/2012ApJ...760..154C} {760, 154}

\bibitem[\protect\citeauthoryear{{Chen}, {Zhang}, {Li}, {Li}  \& {Lin}}{{Chen}
  et~al.}{2020}]{chen2020}
{Chen} Y.-X.,  {Zhang} X.,  {Li} Y.-P.,  {Li} H.,   {Lin} D. N.~C.,  2020,
  \mn@doi [\apj] {10.3847/1538-4357/abaab6}, \href
  {https://ui.adsabs.harvard.edu/abs/2020ApJ...900...44C} {900, 44}

\bibitem[\protect\citeauthoryear{{Cornachione}, {Morgan}, {Millon}, {Bentz},
  {Courbin}, {Bonvin}  \& {Falco}}{{Cornachione}
  et~al.}{2020}]{cornachione2020}
{Cornachione} M.~A.,  {Morgan} C.~W.,  {Millon} M.,  {Bentz} M.~C.,  {Courbin}
  F.,  {Bonvin} V.,   {Falco} E.~E.,  2020, \mn@doi [\apj]
  {10.3847/1538-4357/ab557a}, \href
  {https://ui.adsabs.harvard.edu/abs/2020ApJ...895..125C} {895, 125}

\bibitem[\protect\citeauthoryear{{D'Angelo}, {Kley}  \& {Henning}}{{D'Angelo}
  et~al.}{2003}]{dangelo2003}
{D'Angelo} G.,  {Kley} W.,   {Henning} T.,  2003, \mn@doi [\apj]
  {10.1086/367555}, \href
  {https://ui.adsabs.harvard.edu/abs/2003ApJ...586..540D} {586, 540}

\bibitem[\protect\citeauthoryear{{Davies} \& {Lin}}{{Davies} \&
  {Lin}}{2020}]{davies2020}
{Davies} M.~B.,  {Lin} D. N.~C.,  2020, \mn@doi [\mnras]
  {10.1093/mnras/staa2590}, \href
  {https://ui.adsabs.harvard.edu/abs/2020MNRAS.498.3452D} {498, 3452}

\bibitem[\protect\citeauthoryear{{De Rosa}, {Decarli}, {Walter}, {Fan},
  {Jiang}, {Kurk}, {Pasquali}  \& {Rix}}{{De Rosa} et~al.}{2011}]{derosa2011}
{De Rosa} G.,  {Decarli} R.,  {Walter} F.,  {Fan} X.,  {Jiang} L.,  {Kurk} J.,
  {Pasquali} A.,   {Rix} H.~W.,  2011, \mn@doi [\apj]
  {10.1088/0004-637X/739/2/56}, \href
  {https://ui.adsabs.harvard.edu/abs/2011ApJ...739...56D} {739, 56}

\bibitem[\protect\citeauthoryear{{Dietrich}, {Hamann}, {Appenzeller}  \&
  {Vestergaard}}{{Dietrich} et~al.}{2003}]{dietrich2003}
{Dietrich} M.,  {Hamann} F.,  {Appenzeller} I.,   {Vestergaard} M.,  2003,
  \mn@doi [\apj] {10.1086/378045}, \href
  {https://ui.adsabs.harvard.edu/abs/2003ApJ...596..817D} {596, 817}

\bibitem[\protect\citeauthoryear{{Dobbs-Dixon}, {Li}  \& {Lin}}{{Dobbs-Dixon}
  et~al.}{2007}]{dobbsdixon2007}
{Dobbs-Dixon} I.,  {Li} S.~L.,   {Lin} D.~N.~C.,  2007, \mn@doi [\apj]
  {10.1086/512537}, \href
  {https://ui.adsabs.harvard.edu/abs/2007ApJ...660..791D} {660, 791}

\bibitem[\protect\citeauthoryear{{Duffell} \& {MacFadyen}}{{Duffell} \&
  {MacFadyen}}{2013}]{duffell2013}
{Duffell} P.~C.,  {MacFadyen} A.~I.,  2013, \mn@doi [\apj]
  {10.1088/0004-637X/769/1/41}, \href
  {https://ui.adsabs.harvard.edu/abs/2013ApJ...769...41D} {769, 41}

\bibitem[\protect\citeauthoryear{{Edmunds} \& {Pagel}}{{Edmunds} \&
  {Pagel}}{1978}]{edmunds1978}
{Edmunds} M.~G.,  {Pagel} B.~E.~J.,  1978, \mn@doi [\mnras]
  {10.1093/mnras/185.1.77P}, \href
  {https://ui.adsabs.harvard.edu/abs/1978MNRAS.185P..77E} {185, 77P}

\bibitem[\protect\citeauthoryear{{Elvis}, {Risaliti}  \& {Zamorani}}{{Elvis}
  et~al.}{2002}]{elvis2002}
{Elvis} M.,  {Risaliti} G.,   {Zamorani} G.,  2002, \mn@doi [\apjl]
  {10.1086/339197}, \href
  {https://ui.adsabs.harvard.edu/abs/2002ApJ...565L..75E} {565, L75}

\bibitem[\protect\citeauthoryear{{Fabian} \& {Iwasawa}}{{Fabian} \&
  {Iwasawa}}{1999}]{fabian1999}
{Fabian} A.~C.,  {Iwasawa} K.,  1999, \mn@doi [\mnras]
  {10.1046/j.1365-8711.1999.02404.x}, \href
  {https://ui.adsabs.harvard.edu/abs/1999MNRAS.303L..34F} {303, L34}

\bibitem[\protect\citeauthoryear{{Frank}, {King}  \& {Raine}}{{Frank}
  et~al.}{2002}]{frank2002}
{Frank} J.,  {King} A.,   {Raine} D.~J.,  2002, {Accretion Power in
  Astrophysics: Third Edition}

\bibitem[\protect\citeauthoryear{{Garaud} \& {Lin}}{{Garaud} \&
  {Lin}}{2007}]{garaud2007}
{Garaud} P.,  {Lin} D.~N.~C.,  2007, \mn@doi [\apj] {10.1086/509041}, \href
  {https://ui.adsabs.harvard.edu/abs/2007ApJ...654..606G} {654, 606}

\bibitem[\protect\citeauthoryear{{Goodman}}{{Goodman}}{2003}]{goodman2003}
{Goodman} J.,  2003, \mn@doi [\mnras] {10.1046/j.1365-8711.2003.06241.x}, \href
  {https://ui.adsabs.harvard.edu/abs/2003MNRAS.339..937G} {339, 937}

\bibitem[\protect\citeauthoryear{{Goodman} \& {Tan}}{{Goodman} \&
  {Tan}}{2004}]{goodmantan2004}
{Goodman} J.,  {Tan} J.~C.,  2004, \mn@doi [\apj] {10.1086/386360}, \href
  {https://ui.adsabs.harvard.edu/abs/2004ApJ...608..108G} {608, 108}

\bibitem[\protect\citeauthoryear{{Hamann}, {Korista}, {Ferland}, {Warner}  \&
  {Baldwin}}{{Hamann} et~al.}{2002}]{hamann2002}
{Hamann} F.,  {Korista} K.~T.,  {Ferland} G.~J.,  {Warner} C.,   {Baldwin} J.,
  2002, \mn@doi [\apj] {10.1086/324289}, \href
  {https://ui.adsabs.harvard.edu/abs/2002ApJ...564..592H} {564, 592}

\bibitem[\protect\citeauthoryear{{Hansen}, {Kawaler}  \& {Trimble}}{{Hansen}
  et~al.}{2004}]{hansen2004}
{Hansen} C.~J.,  {Kawaler} S.~D.,   {Trimble} V.,  2004, {Stellar interiors :
  physical principles, structure, and evolution}

\bibitem[\protect\citeauthoryear{Heger \& Woosley}{Heger \&
  Woosley}{2002}]{pair1}
Heger A.,  Woosley S.~E.,  2002, \mn@doi [The Astrophysical Journal]
  {10.1086/338487}, 567, 532

\bibitem[\protect\citeauthoryear{{Henry}, {Edmunds}  \& {K{\"o}ppen}}{{Henry}
  et~al.}{2000}]{henry2000}
{Henry} R.~B.~C.,  {Edmunds} M.~G.,   {K{\"o}ppen} J.,  2000, \mn@doi [\apj]
  {10.1086/309471}, \href
  {https://ui.adsabs.harvard.edu/abs/2000ApJ...541..660H} {541, 660}

\bibitem[\protect\citeauthoryear{{Horne} et~al.,}{{Horne}
  et~al.}{2021}]{horne2021}
{Horne} K.,  et~al., 2021, \mn@doi [\apj] {10.3847/1538-4357/abce60}, \href
  {https://ui.adsabs.harvard.edu/abs/2021ApJ...907...76H} {907, 76}

\bibitem[\protect\citeauthoryear{{Jermyn}, {Dittmann}, {McKernan}, {Ford}  \&
  {Cantiello}}{{Jermyn} et~al.}{2022}]{jermyn2022}
{Jermyn} A.~S.,  {Dittmann} A.~J.,  {McKernan} B.,  {Ford} K.~E.~S.,
  {Cantiello} M.,  2022, \mn@doi [\apj] {10.3847/1538-4357/ac5d40}, \href
  {https://ui.adsabs.harvard.edu/abs/2022ApJ...929..133J} {929, 133}

\bibitem[\protect\citeauthoryear{{Kanagawa}, {Tanaka}, {Muto}, {Tanigawa}  \&
  {Takeuchi}}{{Kanagawa} et~al.}{2015}]{kanagawa2015}
{Kanagawa} K.~D.,  {Tanaka} H.,  {Muto} T.,  {Tanigawa} T.,   {Takeuchi} T.,
  2015, \mn@doi [\mnras] {10.1093/mnras/stv025}, \href
  {https://ui.adsabs.harvard.edu/abs/2015MNRAS.448..994K} {448, 994}

\bibitem[\protect\citeauthoryear{{Kley} \& {Nelson}}{{Kley} \&
  {Nelson}}{2012}]{nelson2012}
{Kley} W.,  {Nelson} R.~P.,  2012, \mn@doi [\araa]
  {10.1146/annurev-astro-081811-125523}, \href
  {https://ui.adsabs.harvard.edu/abs/2012ARA&A..50..211K} {50, 211}

\bibitem[\protect\citeauthoryear{{Kormendy} \& {Ho}}{{Kormendy} \&
  {Ho}}{2013}]{kormendy2013}
{Kormendy} J.,  {Ho} L.~C.,  2013, \mn@doi [\araa]
  {10.1146/annurev-astro-082708-101811}, \href
  {https://ui.adsabs.harvard.edu/abs/2013ARA&A..51..511K} {51, 511}

\bibitem[\protect\citeauthoryear{{Kratter}, {Matzner}  \& {Krumholz}}{{Kratter}
  et~al.}{2008}]{kratter2008}
{Kratter} K.~M.,  {Matzner} C.~D.,   {Krumholz} M.~R.,  2008, \mn@doi [\apj]
  {10.1086/587543}, \href
  {https://ui.adsabs.harvard.edu/abs/2008ApJ...681..375K} {681, 375}

\bibitem[\protect\citeauthoryear{{Lai} et~al.,}{{Lai} et~al.}{2022}]{lai2022}
{Lai} S.,  et~al., 2022, \mn@doi [\mnras] {10.1093/mnras/stac1001}, \href
  {https://ui.adsabs.harvard.edu/abs/2022MNRAS.513.1801L} {513, 1801}

\bibitem[\protect\citeauthoryear{{Langer, N.}, {Norman, C. A.}, {de Koter, A.},
  {Vink, J. S.}, {Cantiello, M.}  \& {Yoon, S.-C.}}{{Langer, N.}
  et~al.}{2007}]{pair4}
{Langer, N.} {Norman, C. A.} {de Koter, A.} {Vink, J. S.} {Cantiello, M.}
  {Yoon, S.-C.} 2007, \mn@doi [A\&A] {10.1051/0004-6361:20078482}, 475, L19

\bibitem[\protect\citeauthoryear{{Li}, {Chen}, {Lin}  \& {Zhang}}{{Li}
  et~al.}{2021}]{lichenlin2021}
{Li} Y.-P.,  {Chen} Y.-X.,  {Lin} D. N.~C.,   {Zhang} X.,  2021, \mn@doi [\apj]
  {10.3847/1538-4357/abc883}, \href
  {https://ui.adsabs.harvard.edu/abs/2021ApJ...906...52L} {906, 52}

\bibitem[\protect\citeauthoryear{{Lin} \& {Papaloizou}}{{Lin} \&
  {Papaloizou}}{1993}]{lin1993}
{Lin} D.~N.~C.,  {Papaloizou} J.~C.~B.,  1993, in {Levy} E.~H.,  {Lunine}
  J.~I.,  eds, Protostars and Planets III. p.~749

\bibitem[\protect\citeauthoryear{{Lin} \& {Pringle}}{{Lin} \&
  {Pringle}}{1987}]{lin1987}
{Lin} D.~N.~C.,  {Pringle} J.~E.,  1987, \mn@doi [\mnras]
  {10.1093/mnras/225.3.607}, \href
  {https://ui.adsabs.harvard.edu/abs/1987MNRAS.225..607L} {225, 607}

\bibitem[\protect\citeauthoryear{{Lin}, {Pringle}  \& {Rees}}{{Lin}
  et~al.}{1988}]{lin1988}
{Lin} D.~N.~C.,  {Pringle} J.~E.,   {Rees} M.~J.,  1988, \mn@doi [\apj]
  {10.1086/166272}, \href
  {https://ui.adsabs.harvard.edu/abs/1988ApJ...328..103L} {328, 103}

\bibitem[\protect\citeauthoryear{{Lynden-Bell}}{{Lynden-Bell}}{1969}]{Lynden-Bell1969}
{Lynden-Bell} D.,  1969, \mn@doi [\nat] {10.1038/223690a0}, \href
  {https://ui.adsabs.harvard.edu/abs/1969Natur.223..690L} {223, 690}

\bibitem[\protect\citeauthoryear{{Lynden-Bell} \& {Pringle}}{{Lynden-Bell} \&
  {Pringle}}{1974}]{lyndenbell1974}
{Lynden-Bell} D.,  {Pringle} J.~E.,  1974, \mn@doi [\mnras]
  {10.1093/mnras/168.3.603}, \href
  {https://ui.adsabs.harvard.edu/abs/1974MNRAS.168..603L} {168, 603}

\bibitem[\protect\citeauthoryear{{Maiolino}, {Juarez}, {Mujica}, {Nagar}  \&
  {Oliva}}{{Maiolino} et~al.}{2003}]{maiolino2003}
{Maiolino} R.,  {Juarez} Y.,  {Mujica} R.,  {Nagar} N.~M.,   {Oliva} E.,  2003,
  \mn@doi [\apjl] {10.1086/379600}, \href
  {https://ui.adsabs.harvard.edu/abs/2003ApJ...596L.155M} {596, L155}

\bibitem[\protect\citeauthoryear{{Marconi}, {Risaliti}, {Gilli}, {Hunt},
  {Maiolino}  \& {Salvati}}{{Marconi} et~al.}{2004}]{marconi2004}
{Marconi} A.,  {Risaliti} G.,  {Gilli} R.,  {Hunt} L.~K.,  {Maiolino} R.,
  {Salvati} M.,  2004, \mn@doi [\mnras] {10.1111/j.1365-2966.2004.07765.x},
  \href {https://ui.adsabs.harvard.edu/abs/2004MNRAS.351..169M} {351, 169}

\bibitem[\protect\citeauthoryear{{Martin} \& {Lubow}}{{Martin} \&
  {Lubow}}{2011}]{martin2011}
{Martin} R.~G.,  {Lubow} S.~H.,  2011, \mn@doi [\apjl]
  {10.1088/2041-8205/740/1/L6}, \href
  {https://ui.adsabs.harvard.edu/abs/2011ApJ...740L...6M} {740, L6}

\bibitem[\protect\citeauthoryear{{Meynet} \& {Maeder}}{{Meynet} \&
  {Maeder}}{2002}]{meynet2002}
{Meynet} G.,  {Maeder} A.,  2002, \mn@doi [\aap] {10.1051/0004-6361:20011554},
  \href {https://ui.adsabs.harvard.edu/abs/2002A&A...381L..25M} {381, L25}

\bibitem[\protect\citeauthoryear{{Miller} \& {Scalo}}{{Miller} \&
  {Scalo}}{1979}]{miller1979}
{Miller} G.~E.,  {Scalo} J.~M.,  1979, \mn@doi [\apjs] {10.1086/190629}, \href
  {https://ui.adsabs.harvard.edu/abs/1979ApJS...41..513M} {41, 513}

\bibitem[\protect\citeauthoryear{{Morgan}, {Hyer}, {Bonvin}, {Mosquera},
  {Cornachione}, {Courbin}, {Kochanek}  \& {Falco}}{{Morgan}
  et~al.}{2018}]{morgan2018}
{Morgan} C.~W.,  {Hyer} G.~E.,  {Bonvin} V.,  {Mosquera} A.~M.,  {Cornachione}
  M.,  {Courbin} F.,  {Kochanek} C.~S.,   {Falco} E.~E.,  2018, \mn@doi [\apj]
  {10.3847/1538-4357/aaed3e}, \href
  {https://ui.adsabs.harvard.edu/abs/2018ApJ...869..106M} {869, 106}

\bibitem[\protect\citeauthoryear{{Nagao}, {Marconi}  \& {Maiolino}}{{Nagao}
  et~al.}{2006a}]{nagao2006a}
{Nagao} T.,  {Marconi} A.,   {Maiolino} R.,  2006a, \mn@doi [\aap]
  {10.1051/0004-6361:20054024}, \href
  {https://ui.adsabs.harvard.edu/abs/2006A&A...447..157N} {447, 157}

\bibitem[\protect\citeauthoryear{{Nagao}, {Maiolino}  \& {Marconi}}{{Nagao}
  et~al.}{2006b}]{nagao2006b}
{Nagao} T.,  {Maiolino} R.,   {Marconi} A.,  2006b, \mn@doi [\aap]
  {10.1051/0004-6361:20054127}, \href
  {https://ui.adsabs.harvard.edu/abs/2006A&A...447..863N} {447, 863}

\bibitem[\protect\citeauthoryear{{Osterbrock} \& {Ferland}}{{Osterbrock} \&
  {Ferland}}{2006}]{osterbrock2006}
{Osterbrock} D.~E.,  {Ferland} G.~J.,  2006, {Astrophysics of gaseous nebulae
  and active galactic nuclei}

\bibitem[\protect\citeauthoryear{{Osterbrock} \& {Shuder}}{{Osterbrock} \&
  {Shuder}}{1982}]{osterbrock1982}
{Osterbrock} D.~E.,  {Shuder} J.~M.,  1982, \mn@doi [\apjs] {10.1086/190793},
  \href {https://ui.adsabs.harvard.edu/abs/1982ApJS...49..149O} {49, 149}

\bibitem[\protect\citeauthoryear{{Owocki} \& {Shaviv}}{{Owocki} \&
  {Shaviv}}{2012}]{owocki2012}
{Owocki} S.~P.,  {Shaviv} N.~J.,  2012, in {Davidson} K.,  {Humphreys} R.~M.,
  eds,  Astrophysics and Space Science Library Vol. 384, Eta Carinae and the
  Supernova Impostors. p.~275, \mn@doi{10.1007/978-1-4614-2275-4_12}

\bibitem[\protect\citeauthoryear{{Owocki}, {Gayley}  \& {Shaviv}}{{Owocki}
  et~al.}{2004}]{shaviv}
{Owocki} S.~P.,  {Gayley} K.~G.,   {Shaviv} N.~J.,  2004, \mn@doi [\apj]
  {10.1086/424910}, \href
  {https://ui.adsabs.harvard.edu/abs/2004ApJ...616..525O} {616, 525}

\bibitem[\protect\citeauthoryear{{Paczynski}}{{Paczynski}}{1978}]{paczynski1978}
{Paczynski} B.,  1978, \actaa, \href
  {https://ui.adsabs.harvard.edu/abs/1978AcA....28...91P} {28, 91}

\bibitem[\protect\citeauthoryear{{Paxton}, {Bildsten}, {Dotter}, {Herwig},
  {Lesaffre}  \& {Timmes}}{{Paxton} et~al.}{2011}]{mesa1}
{Paxton} B.,  {Bildsten} L.,  {Dotter} A.,  {Herwig} F.,  {Lesaffre} P.,
  {Timmes} F.,  2011, \mn@doi [\apjs] {10.1088/0067-0049/192/1/3}, \href
  {https://ui.adsabs.harvard.edu/abs/2011ApJS..192....3P} {192, 3}

\bibitem[\protect\citeauthoryear{{Paxton} et~al.,}{{Paxton}
  et~al.}{2013}]{mesa2}
{Paxton} B.,  et~al., 2013, \mn@doi [\apjs] {10.1088/0067-0049/208/1/4}, \href
  {https://ui.adsabs.harvard.edu/abs/2013ApJS..208....4P} {208, 4}

\bibitem[\protect\citeauthoryear{{Paxton} et~al.,}{{Paxton}
  et~al.}{2015}]{mesa3}
{Paxton} B.,  et~al., 2015, \mn@doi [\apjs] {10.1088/0067-0049/220/1/15}, \href
  {https://ui.adsabs.harvard.edu/abs/2015ApJS..220...15P} {220, 15}

\bibitem[\protect\citeauthoryear{{Paxton} et~al.,}{{Paxton}
  et~al.}{2018}]{mesa4}
{Paxton} B.,  et~al., 2018, \mn@doi [\apjs] {10.3847/1538-4365/aaa5a8}, \href
  {https://ui.adsabs.harvard.edu/abs/2018ApJS..234...34P} {234, 34}

\bibitem[\protect\citeauthoryear{{Paxton} et~al.,}{{Paxton}
  et~al.}{2019}]{mesa5}
{Paxton} B.,  et~al., 2019, \mn@doi [\apjs] {10.3847/1538-4365/ab2241}, \href
  {https://ui.adsabs.harvard.edu/abs/2019ApJS..243...10P} {243, 10}

\bibitem[\protect\citeauthoryear{{Peimbert}, {Peimbert}  \&
  {Luridiana}}{{Peimbert} et~al.}{2016}]{peimbert2016}
{Peimbert} A.,  {Peimbert} M.,   {Luridiana} V.,  2016, \rmxaa, \href
  {https://ui.adsabs.harvard.edu/abs/2016RMxAA..52..419P} {52, 419}

\bibitem[\protect\citeauthoryear{{Pooley}, {Blackburne}, {Rappaport}  \&
  {Schechter}}{{Pooley} et~al.}{2007}]{pooley2007}
{Pooley} D.,  {Blackburne} J.~A.,  {Rappaport} S.,   {Schechter} P.~L.,  2007,
  \mn@doi [\apj] {10.1086/512115}, \href
  {https://ui.adsabs.harvard.edu/abs/2007ApJ...661...19P} {661, 19}

\bibitem[\protect\citeauthoryear{{Pringle}}{{Pringle}}{1981}]{pringle1981}
{Pringle} J.~E.,  1981, \mn@doi [\araa] {10.1146/annurev.aa.19.090181.001033},
  \href {https://ui.adsabs.harvard.edu/abs/1981ARA&A..19..137P} {19, 137}

\bibitem[\protect\citeauthoryear{{Rafikov}}{{Rafikov}}{2015}]{rafikov2015}
{Rafikov} R.~R.,  2015, \mn@doi [\apj] {10.1088/0004-637X/804/1/62}, \href
  {https://ui.adsabs.harvard.edu/abs/2015ApJ...804...62R} {804, 62}

\bibitem[\protect\citeauthoryear{{Raimundo} \& {Fabian}}{{Raimundo} \&
  {Fabian}}{2009}]{raim}
{Raimundo} S.~I.,  {Fabian} A.~C.,  2009, \mn@doi [\mnras]
  {10.1111/j.1365-2966.2009.14796.x}, \href
  {https://ui.adsabs.harvard.edu/abs/2009MNRAS.396.1217R} {396, 1217}

\bibitem[\protect\citeauthoryear{{Rees}}{{Rees}}{1984}]{rees1984}
{Rees} M.~J.,  1984, \mn@doi [\araa] {10.1146/annurev.aa.22.090184.002351},
  \href {https://ui.adsabs.harvard.edu/abs/1984ARA&A..22..471R} {22, 471}

\bibitem[\protect\citeauthoryear{{Rosenthal}, {Chiang}, {Ginzburg}  \&
  {Murray-Clay}}{{Rosenthal} et~al.}{2020}]{rosenthal2020}
{Rosenthal} M.~M.,  {Chiang} E.~I.,  {Ginzburg} S.,   {Murray-Clay} R.~A.,
  2020, \mn@doi [\mnras] {10.1093/mnras/staa1721}, \href
  {https://ui.adsabs.harvard.edu/abs/2020MNRAS.498.2054R} {498, 2054}

\bibitem[\protect\citeauthoryear{{Safronov}}{{Safronov}}{1960}]{safronov1960}
{Safronov} V.~S.,  1960, Annales d'Astrophysique, \href
  {https://ui.adsabs.harvard.edu/abs/1960AnAp...23..979S} {23, 979}

\bibitem[\protect\citeauthoryear{{Sanders}, {Phinney}, {Neugebauer}, {Soifer}
  \& {Matthews}}{{Sanders} et~al.}{1989}]{sanders1989}
{Sanders} D.~B.,  {Phinney} E.~S.,  {Neugebauer} G.,  {Soifer} B.~T.,
  {Matthews} K.,  1989, \mn@doi [\apj] {10.1086/168094}, \href
  {https://ui.adsabs.harvard.edu/abs/1989ApJ...347...29S} {347, 29}

\bibitem[\protect\citeauthoryear{{Shakura} \& {Sunyaev}}{{Shakura} \&
  {Sunyaev}}{1973}]{shakura1973}
{Shakura} N.~I.,  {Sunyaev} R.~A.,  1973, \aap, \href
  {https://ui.adsabs.harvard.edu/abs/1973A&A....24..337S} {24, 337}

\bibitem[\protect\citeauthoryear{{Shang}, {Wills}, {Wills}  \&
  {Brotherton}}{{Shang} et~al.}{2007}]{shang2007}
{Shang} Z.,  {Wills} B.~J.,  {Wills} D.,   {Brotherton} M.~S.,  2007, \mn@doi
  [\aj] {10.1086/518505}, \href
  {https://ui.adsabs.harvard.edu/abs/2007AJ....134..294S} {134, 294}

\bibitem[\protect\citeauthoryear{{Shankar}, {Salucci}, {Granato}, {De Zotti}
  \& {Danese}}{{Shankar} et~al.}{2004}]{shankar2004}
{Shankar} F.,  {Salucci} P.,  {Granato} G.~L.,  {De Zotti} G.,   {Danese} L.,
  2004, \mn@doi [\mnras] {10.1111/j.1365-2966.2004.08261.x}, \href
  {https://ui.adsabs.harvard.edu/abs/2004MNRAS.354.1020S} {354, 1020}

\bibitem[\protect\citeauthoryear{{Shankar}, {Weinberg}  \&
  {Miralda-Escud{\'e}}}{{Shankar} et~al.}{2009}]{shankar2009}
{Shankar} F.,  {Weinberg} D.~H.,   {Miralda-Escud{\'e}} J.,  2009, \mn@doi
  [\apj] {10.1088/0004-637X/690/1/20}, \href
  {https://ui.adsabs.harvard.edu/abs/2009ApJ...690...20S} {690, 20}

\bibitem[\protect\citeauthoryear{{Soltan}}{{Soltan}}{1982}]{soltan1982}
{Soltan} A.,  1982, \mn@doi [\mnras] {10.1093/mnras/200.1.115}, \href
  {https://ui.adsabs.harvard.edu/abs/1982MNRAS.200..115S} {200, 115}

\bibitem[\protect\citeauthoryear{{Spera} \& {Mapelli}}{{Spera} \&
  {Mapelli}}{2017}]{spera2017}
{Spera} M.,  {Mapelli} M.,  2017, \mn@doi [\mnras] {10.1093/mnras/stx1576},
  \href {https://ui.adsabs.harvard.edu/abs/2017MNRAS.470.4739S} {470, 4739}

\bibitem[\protect\citeauthoryear{{Starkey}, {Huang}, {Horne}  \&
  {Lin}}{{Starkey} et~al.}{2022}]{starkey2022}
{Starkey} D.~A.,  {Huang} J.,  {Horne} K.,   {Lin} D. N.~C.,  2022, \mn@doi
  [\mnras] {10.1093/mnras/stac3579}, \href
  {https://ui.adsabs.harvard.edu/abs/2022MNRAS.tmp.3495S} {}

\bibitem[\protect\citeauthoryear{{Sukhbold}, {Ertl}, {Woosley}, {Brown}  \&
  {Janka}}{{Sukhbold} et~al.}{2016}]{sukhbold2016}
{Sukhbold} T.,  {Ertl} T.,  {Woosley} S.~E.,  {Brown} J.~M.,   {Janka} H.~T.,
  2016, \mn@doi [\apj] {10.3847/0004-637X/821/1/38}, \href
  {https://ui.adsabs.harvard.edu/abs/2016ApJ...821...38S} {821, 38}

\bibitem[\protect\citeauthoryear{{Syer}, {Clarke}  \& {Rees}}{{Syer}
  et~al.}{1991}]{syer1991}
{Syer} D.,  {Clarke} C.~J.,   {Rees} M.~J.,  1991, \mn@doi [\mnras]
  {10.1093/mnras/250.3.505}, \href
  {https://ui.adsabs.harvard.edu/abs/1991MNRAS.250..505S} {250, 505}

\bibitem[\protect\citeauthoryear{{Talbot} \& {Arnett}}{{Talbot} \&
  {Arnett}}{1974}]{talbot1974}
{Talbot} R. J.~J.,  {Arnett} D.~W.,  1974, \mn@doi [\apj] {10.1086/152918},
  \href {https://ui.adsabs.harvard.edu/abs/1974ApJ...190..605T} {190, 605}

\bibitem[\protect\citeauthoryear{{Tanigawa} \& {Tanaka}}{{Tanigawa} \&
  {Tanaka}}{2016}]{tanigawa2016}
{Tanigawa} T.,  {Tanaka} H.,  2016, \mn@doi [\apj]
  {10.3847/0004-637X/823/1/48}, \href
  {https://ui.adsabs.harvard.edu/abs/2016ApJ...823...48T} {823, 48}

\bibitem[\protect\citeauthoryear{{Temple}, {Ferland}, {Rankine}, {Chatzikos}
  \& {Hewett}}{{Temple} et~al.}{2021}]{temple2021}
{Temple} M.~J.,  {Ferland} G.~J.,  {Rankine} A.~L.,  {Chatzikos} M.,   {Hewett}
  P.~C.,  2021, \mn@doi [\mnras] {10.1093/mnras/stab1610}, \href
  {https://ui.adsabs.harvard.edu/abs/2021MNRAS.505.3247T} {505, 3247}

\bibitem[\protect\citeauthoryear{{Terao}, {Nagao}, {Onishi}, {Matsuoka},
  {Akiyama}, {Matsuoka}  \& {Yamashita}}{{Terao} et~al.}{2022}]{terao2022}
{Terao} K.,  {Nagao} T.,  {Onishi} K.,  {Matsuoka} K.,  {Akiyama} M.,
  {Matsuoka} Y.,   {Yamashita} T.,  2022, \mn@doi [\apj]
  {10.3847/1538-4357/ac5b71}, \href
  {https://ui.adsabs.harvard.edu/abs/2022ApJ...929...51T} {929, 51}

\bibitem[\protect\citeauthoryear{{Thompson}, {Quataert}  \&
  {Murray}}{{Thompson} et~al.}{2005}]{thompson2005}
{Thompson} T.~A.,  {Quataert} E.,   {Murray} N.,  2005, \mn@doi [\apj]
  {10.1086/431923}, \href
  {https://ui.adsabs.harvard.edu/abs/2005ApJ...630..167T} {630, 167}

\bibitem[\protect\citeauthoryear{{Thuan}, {Izotov}  \& {Lipovetsky}}{{Thuan}
  et~al.}{1995}]{thuan1995}
{Thuan} T.~X.,  {Izotov} Y.~I.,   {Lipovetsky} V.~A.,  1995, \mn@doi [\apj]
  {10.1086/175676}, \href
  {https://ui.adsabs.harvard.edu/abs/1995ApJ...445..108T} {445, 108}

\bibitem[\protect\citeauthoryear{{Tomkin} \& {Lambert}}{{Tomkin} \&
  {Lambert}}{1984}]{tomkin1984}
{Tomkin} J.,  {Lambert} D.~L.,  1984, \mn@doi [\apj] {10.1086/161885}, \href
  {https://ui.adsabs.harvard.edu/abs/1984ApJ...279..220T} {279, 220}

\bibitem[\protect\citeauthoryear{{Toomre}}{{Toomre}}{1964}]{toomre1964}
{Toomre} A.,  1964, \mn@doi [\apj] {10.1086/147861}, \href
  {https://ui.adsabs.harvard.edu/abs/1964ApJ...139.1217T} {139, 1217}

\bibitem[\protect\citeauthoryear{{Valerdi}, {Peimbert}  \&
  {Peimbert}}{{Valerdi} et~al.}{2021}]{valerdi2021}
{Valerdi} M.,  {Peimbert} A.,   {Peimbert} M.,  2021, \mn@doi [\mnras]
  {10.1093/mnras/stab1543}, \href
  {https://ui.adsabs.harvard.edu/abs/2021MNRAS.505.3624V} {505, 3624}

\bibitem[\protect\citeauthoryear{{Wang} et~al.,}{{Wang}
  et~al.}{2022}]{wang2022}
{Wang} S.,  et~al., 2022, \mn@doi [\apj] {10.3847/1538-4357/ac3a69}, \href
  {https://ui.adsabs.harvard.edu/abs/2022ApJ...925..121W} {925, 121}

\bibitem[\protect\citeauthoryear{{Woosley}}{{Woosley}}{2017}]{woosley2017}
{Woosley} S.~E.,  2017, \mn@doi [\apj] {10.3847/1538-4357/836/2/244}, \href
  {https://ui.adsabs.harvard.edu/abs/2017ApJ...836..244W} {836, 244}

\bibitem[\protect\citeauthoryear{{Xu}, {Bian}, {Shen}, {Zuo}, {Fan}  \&
  {Zhu}}{{Xu} et~al.}{2018}]{xu2018}
{Xu} F.,  {Bian} F.,  {Shen} Y.,  {Zuo} W.,  {Fan} X.,   {Zhu} Z.,  2018,
  \mn@doi [\mnras] {10.1093/mnras/sty1763}, \href
  {https://ui.adsabs.harvard.edu/abs/2018MNRAS.480..345X} {480, 345}

\bibitem[\protect\citeauthoryear{{Yu} \& {Tremaine}}{{Yu} \&
  {Tremaine}}{2002}]{yu2002}
{Yu} Q.,  {Tremaine} S.,  2002, \mn@doi [\mnras]
  {10.1046/j.1365-8711.2002.05532.x}, \href
  {https://ui.adsabs.harvard.edu/abs/2002MNRAS.335..965Y} {335, 965}

\bibitem[\protect\citeauthoryear{{Zhu}, {Hartmann}, {Gammie}, {Book}, {Simon}
  \& {Engelhard}}{{Zhu} et~al.}{2010a}]{zhu2010a}
{Zhu} Z.,  {Hartmann} L.,  {Gammie} C.~F.,  {Book} L.~G.,  {Simon} J.~B.,
  {Engelhard} E.,  2010a, \mn@doi [\apj] {10.1088/0004-637X/713/2/1134}, \href
  {https://ui.adsabs.harvard.edu/abs/2010ApJ...713.1134Z} {713, 1134}

\bibitem[\protect\citeauthoryear{{Zhu}, {Hartmann}  \& {Gammie}}{{Zhu}
  et~al.}{2010b}]{zhu2010b}
{Zhu} Z.,  {Hartmann} L.,   {Gammie} C.,  2010b, \mn@doi [\apj]
  {10.1088/0004-637X/713/2/1143}, \href
  {https://ui.adsabs.harvard.edu/abs/2010ApJ...713.1143Z} {713, 1143}

\makeatother
\end{thebibliography}


\appendix

\bsp	
\label{lastpage}

\end{document}